\documentclass[fleqn,usenatbib,useAMS]{mnras}

\usepackage{graphicx}	
\usepackage{amsmath}	
\usepackage{amssymb}	
\usepackage{multicol}        
\usepackage{bm}		
\usepackage{pdflscape}	
\usepackage{color}
\usepackage{multirow}
\usepackage{lineno}
\usepackage{adjustbox}
\usepackage{hyperref}
\usepackage[hyphenbreaks]{breakurl}

\usepackage[T1]{fontenc}
\usepackage{ae,aecompl}

\title{Classifications of Fermi-LAT unassociated sources in multiple machine learning methods}

\author[K.R. Zhu et al.]{
K.R. Zhu $^{1}$
J.M. Chen $^{1}$
Y.G. Zheng $^{2}$
L. Zhang ,$^{1}$\thanks{E-mail: lizhang@ynu.edu.cn}
\\
$^{1}$Department of Astronomy, School of Physics and Astronomy, Yunnan University, Kunming, Yunnan, 650091,  People's Republic of China\\
$^{2}$Department of Physics, Yunnan Normal University, Kunming, Yunnan, 650092, People's Republic of China\
}

\date{Accepted XXX. Received YYY; in original form ZZZ}

\pubyear{2023}

\begin{document}
\label{firstpage}
\pagerange{\pageref{firstpage}--\pageref{lastpage}}
\maketitle
\begin{abstract}
The classifications of Fermi-LAT unassociated sources are studied using multiple machine learning (ML) methods. The update data from 4FGL-DR3 are divided into high Galactic latitude (HGL, Galactic latitude $|b|>10^\circ$) and low Galactic latitude (LGL, $|b|\le10^\circ$) regions. In the HGL region, a voting ensemble of four binary ML classifiers achieves a 91$\%$ balanced accuracy. In the LGL region, an additional Bayesian-Gaussian (BG) model with three parameters is introduced to eliminate abnormal soft spectrum AGNs from the training set and ML-identified AGN candidates, a voting ensemble of four ternary ML algorithms reach an 81$\%$ balanced accuracy. And then, a catalog of Fermi-LAT all-sky unassociated sources is constructed. Our classification results show that (i) there are 1037 AGN candidates and 88 pulsar candidates with a balanced accuracy of $0.918 \pm 0.029$ in HGL region, which are consistent with those given in previous all-sky ML approaches; and (ii) there are 290 AGN-like candidates, 135 pulsar-like candidates, and 742 other-like candidates with a balanced accuracy of $0.815 \pm 0.027$ in the LGL region, which are different from those in previous all-sky ML approaches. Additionally, different training sets and class weights were tested for their impact on classifier accuracy and predicted results. The findings suggest that while different training approaches can yield similar model accuracy, the predicted numbers across different categories can vary significantly. Thus, reliable evaluation of the predicted results is deemed crucial in the ML approach for Fermi-LAT unassociated sources.
\end{abstract}




\section{Introduction}\label{sec:intro}

Early gamma-ray source catalogs, such as Celestial Observation Satellite (COS-B) source catalogs (e.g. \citealt{1981RSPTA.301..519H,1987ICRC....1...88P}) and Compton Gamma Ray Observatory (CGRO) source catalogs (e.g. \citealt{1994ApJS...94..551F,1995ApJS..101..259T,1999ApJS..123...79H}), only included a small number of sources, and most of them were not identified or associated in other wavelength bands, label as unassociated sources. The detection capability of GeV gamma-ray sources was greatly enhanced with the launch of the Fermi Gamma-ray Space Telescope in 2008 \citep{2009ApJ...697.1071A}. The Fermi-LAT collaboration regularly releases catalogs of Fermi-LAT GeV gamma-ray sources (FGL). With increasing exposure time and improved gamma-ray background modeling, a large number of gamma-ray sources have been detected. Recent releases of the Fermi-LAT source catalogs, such as 3FGL \citep{2015ApJS..218...23A}, 4FGL-DR1 \citep{2020ApJS..247...33A}, 4FGL-DR2 \citep{2020arXiv200511208B}, and 4FGL-DR3 \citep{2022ApJS..260...53A}, contain thousands of gamma-ray point sources. Alongside the detection of a large number of gamma-ray sources, a significant number of dark, non-variable unassociated sources have been discovered.

Recently, the Fermi-LAT collaboration released a new version of the Fermi Large Area Telescope (FGL) catalog, known as 4FGL-DR3 \citep{2022ApJS..260...53A}. This updated catalog comprises 6658 point sources, making it the largest gamma-ray source catalog to date. However, approximately one-third of these sources remain unassociated with known counterparts. Based on a Galactic latitude cut at $|b|=10^{\circ}$, \cite{2022ApJS..260...53A} divides the sky into high Galactic latitude (HGL) and low Galactic latitude (LGL) regions. The parameter distributions of unassociated sources were analysed in both regions, indicating that HGL unassociated sources are likely dominated by blazar-like objects, while the composition of sources in the LGL region may be more complex. Based on the assumption of a uniform distribution of blazars in the entire sky and considering the background contamination at low latitudes, an estimation of detectable blazars in the low-latitude region was performed. The number of blazars in the low-latitude region is limited to $340\pm20$. Additionally, It was also pointed out that the distribution of spectral indices for low-latitude blazars shows anomalies, with an excess of $75\pm4$ soft-spectrum sources, which may be attributed to contamination from the Galactic component.

The classification of unassociated sources and finding their multi-wavelength counterparts are important scientific goals. It has significant implications for understanding high-energy radiation mechanisms, the origin of cosmic rays, and other astrophysical phenomena. Unfortunately, most of the unassociated sources are faint and exhibit weak variability, with their significance often close to the detection threshold. Moreover, unassociated sources are predominantly concentrated in LGL regions, where the presence of strong diffuse gamma-ray background and high source density near the Galactic plane makes their detection and identification more challenging.

Statistical methods (e.g. \citealt{2012ApJ...753...83A}) or multi-band characterization (e.g. \citealt{2018MNRAS.475..942F,2021ApJ...908..177K}) have been used for the classification of Fermi-LAT unassociated sources. In recent years, machine learning (ML) has achieved success in the field of big data mining and analysis, and it has been widely applied to astronomical data \citep{2019arXiv190407248B}.
ML can be divided into supervised learning and unsupervised learning. Classification mainly refers to the application of supervised ML (referred to as ML below). ML methods have been widely applied to the classification of Fermi-LAT unassociated sources \citep{2012MNRAS.424L..64M,2014ApJ...782...41D,2016ApJ...820....8S,2016ApJ...825...69M,2017A&A...602A..86L,2020MNRAS.tmp..163L,2021RAA....21...15Z,2021MNRAS.507.4061F,2021MNRAS.505.5853G,2021JHEAp..29...40C,2022A&A...660A..87B,2022MNRAS.515.1807C,2023MNRAS.521.6195M}, achieving high levels of accuracy in the training set and test set (e.g., >95$\%$, see \citealt{2022A&A...660A..87B,2022MNRAS.515.1807C}).

However, there are still some issues that need to be addressed. The sample representativeness is one of the fundamental assumptions in ML applications \citep{bishop2006pattern}. It requires that the training dataset and predicted samples are sampled independently and identically from the overall data distribution, capable of representing the features and patterns of the entire dataset. Only by fulfilling this assumption can we ensure that ML models have good accuracy and generalization ability when making predictions.

In the task of ML classification Fermi-LAT unassociated sources, the sources used for training the models are primarily those with high significance and strong variability. However, it is questionable whether such models can be effectively applied to predict the nature of dark sources \citep{2021RAA....21...15Z}. Moreover, The associated sources are mainly dominated by HGL active galactic nuclei (AGNs), while most of the unassociated sources are located in LGL regions, which may be dominated by the Galactic population. Due to the strong diffuse gamma-ray background in the neighbouring region of the Galactic plane and the distribution difference of Galactic and extragalactic sources between the HGL and LGL regions, it remains uncertain whether the all-sky models trained mainly by HGL sources are suitable for classifying LGL sources.

In addition, previous attempts have primarily focused on optimizing the performance of models on training and test sets, but lacked the necessary assessment of the credibility of prediction results. For instance, the high density of LGL AGN-like candidates resulting from LGL unassociated sources, as well as the different distributions between predicted candidates and associated samples in the same parameter space.

In this work, we established classification models in the HGL and LGL regions, respectively. Different feature sets and classification strategies were chosen based on different datasets. The models were trained and optimized, and the validity of the prediction results was evaluated when classification results were obtained. In particular, in the LGL region, we developed a Bayesian-Gaussian (BG) model that incorporates spectral index to eliminate the excess of soft spectral AGNs in the training samples,  and to reassess the AGN-like candidates obtained from ML classification. Combining the classification results from both high and low-latitude regions, we established a catalog of unassociated sources across the all sky. At the all-sky scale, we conducted an analysis of the plausibility of the classification results and compared the differences between the models in the HGL and LGL regions.

The structure of the paper is as follows.
Section 2 describes the process of data collection, selection, and preprocessing.
Section 3 provides a brief overview of ML classifiers, where training, optimization, and ensemble methods are introduced.
Section 4 provides a detailed description of the training, testing process, and classification results of the LGL supervised learning classifier.
Section 5 presents the establishment of the low-latitude classification model. Specifically, Section 5.1 describes the development and utilization of the BG model, while Section 5.2 discuss the training, optimization, and classification results of the ML model.
Section 6 combines the results obtained in the previous sections to construct a catalog of all-sky  unassociated sources, and provides the distribution of candidates at the all-sky scale.
The conclusion and discussion are presented in Section 7.

To ensure the reproducibility of our work, we fixed the random seed to ``123'' in the code involving random processes.

\begin{table*}
\centering
\caption{The feature parameters used for classification}\label{Tab1}
\resizebox{\textwidth}{!}{
\begin{tabular}{cccccc}
\hline \hline
\multirow{2}{*}{Label}&\multirow{2}{*}{Feature}& \multirow{2}{*}{Symbol	}&\multirow{2}{*}{Unit}&	 \multirow{2}{*}{Description}	&\multirow{2}{*}{Log10}\\
&&&&&\\
\normalsize(1) & \normalsize(2) & \normalsize(3) &\normalsize(4) & \normalsize(5) &  \normalsize(6) \\
\hline
1&	GLON	&	$l$	&	deg	&	Galactic longitude	&	N	\\
2&	GLAT	&	$b$	&	deg	&	Galactic latitude	&	N	\\
3&	PL$\_$Index	&	$ \Gamma $	&		&	Spectral index when fitting with PowerLaw spectrum	&	N	\\
4&	PL$\_$Flux$\_$Density	&	$N_0 (\rm{PL})$	&	MeV$^{-1}$ cm$^{-2}$ s$^{-1}$	&	Differential flux at pivot energy when fitting with PowerLaw spectrum	&	 Y	\\
5&	LP$\_$Index	&	$\alpha$	&		&	Spectral index when fitting with LogParabola spectrum	&	 N	\\
6&	LP$\_$beta	&	$\beta$	&		&	Spectral curvature when fitting with LogParabola spectrum	&	N	\\
7&	LP$\_$Flux$\_$Density	&	$N_0 (\rm{LP})$	&	MeV$^{-1}$ cm$^{-2}$ s$^{-1}$	&	Differential flux at pivot energy when fitting with LogParabola spectrum	&	Y	\\
8&	LP$\_$SigCurv	&	$TS_{\rm{LP}}$	&		&	Significance of the fit improvement between PowerLaw and LogParabola 	&	Y	\\
9&	PLEC$\_$IndexS	&	$\Gamma_s$	&		&	Spectral index at pivot energy when fitting with PLSuperExpCutoff4 spectrum	&	N	\\
10&	PLEC$\_$ExpfactorS	&	$\beta_s$	&		&	Spectral curvature at pivot energy when fitting with PLSuperExpCutoff4 spectrum	&	N	\\
11&	PLEC$\_$Flux$\_$Density	&	$N_0 (\rm{PLEC})$	&	MeV$^{-1}$ cm$^{-2}$ s$^{-1}$	&	Differential flux at pivot energy when fitting with PLSuperExpCutoff4 spectrum	&	Y	\\
12&	PLEC$\_$SigCurv	&	$TS_{\rm{PLEC}}$	&		&	Significance of the fit improvement between PowerLaw and PLSuperExpCutoff4 	&	Y	\\
\multirow{2}{*}{13}&\multirow{2}{*}{	Variability$\_$Index}&\multirow{2}{*}{$Var$}&	&	Sum of difference between the flux in each time interval and the average flux&	\multirow{2}{*}{Y}\\
&&&& over the full catalog interval&\\
14&	Frac$\_$Variability	&	$F_{var}$	&		&	Fractional variability computed from the fluxes in each year	&	N	\\
15&	Signif$\_$Avg	&	$Sig$	&		&	Source significance over the 100 MeV to 1 TeV	&	Y	\\
16&	Npred	&	$N_{pre}$	&		&	Predicted number of events in the model	&	Y	\\
17&	Flux1000	&	$F_{1000}$	&	cm$^{-2}$ s$^{-1}$	&	Integral photon flux from 1 to 100 GeV	&	Y	\\
18&	Energy$\_$Flux100	&	$E_{100}$	&	erg
 cm$^{-2}$ s$^{-1}$	&	Energy flux from 100 MeV to 100 GeV obtained by spectral fitting	&	Y	\\
19-26&	nuFnu$\_$Band	&	$\nu F_{\nu} (i)$	&	erg cm$^{-2}$ s$^{-1}$	&	Spectral energy distribution over 8 Fermi-LAT band	&	Y	\\
27-33&	Hardness ratio	&	$hr (i, j)$	&		&	Hardness ratios of 8 Fermi-LAT band, as defined in Equation \ref{eq1}	&	N	\\
34-39&	concavity$\_$convexity	&	$H (i, j, k)$	&		&	Parameters of spectral concavity and convexity of 8 Fermi-LAT band, as defined in Equation \ref{eq2}	&	N	\\
\hline
\end{tabular}}\\
{\footnotesize{{Note: Column (1)-(4) are the feature number, name, symbol, and unit used for classification; Column (5) describes of the physical meaning of the feature; Column (6) indicates whether the feature has been logarithmically transformed. }} }
\end{table*}

\section{Dataset preparation and preprocessing} \label{sec:dataset}

The Fermi-LAT Collaboration published the Fourth Fermi-LAT Gamma-ray Source Catalog (4FGL) in 2020, covering the results of Fermi-LAT's sky survey observations from 2008 to 2016, spanning a period of eight years  \citep{2020ApJS..247...33A}. Subsequently, the Fermi-LAT Collaboration updates the 4FGL catalog every two years, incorporating an additional two years of observational data. The latest version of the Fourth Fermi-LAT Gamma-ray Source Catalog is the third release (4FGL-DR3), published in 2022, encompassing GeV observational data from 2008 to 2020 \citep{2022ApJS..260...53A}. It contained the largest catalog of gamma-ray sources in the GeV energy range to date.

4FGL-DR3 \footnote{gll$\_$psc$\_$v30.fit, see \url{https://fermi.gsfc.nasa.gov/ssc/data/access/lat/12yr_catalog/gll_psc_v30.fit}. Please note that this work is based on the v30 version. The latest version v31 has made modifications to some keywords and incorrect TeV associations, but it has not changed the source classes and parameters used in this paper.} contains 6658 point sources. Among them, 4367 sources have been identified or associated with counterparts in other wavebands and are classified into 22 subclasses. 134 sources with LGL are weakly associated with X-ray counterparts but their nature is unknown. Additionally, 2157 sources have not been found to have counterparts in other wavebands and are referred to as unassociated sources. Based on the distribution of GeV gamma-ray sources, the 6,658 point sources can be categorized into 4 major classes:

\begin{itemize}
  \item [1.]
AGN-like class, which includes different types of blazars (FSRQ, fsrq, BLL, bll, BCU, bcu\footnote{The class label in the 4FGL-DR3 table.}) and non-blazar AGN (RDG, rdg, AGN, agn, css, NLSY1, nlsy1, sey), characterized by significant flux variability. There are a total of 3813 sources in the AGN class, with 3406 at high Galactic latitude area ($|b| \geq 10^{\circ}$) and 407 at low Galactic latitude  area ($|b|<10^{\circ}$). They are labeled as ``agn" in the context.
  \item [2.]
Pulsar-like class, which includes millisecond pulsars (MSP, msp) and young pulsars (PSR, psr), characterized by curved spectra and spectral cutoffs in the GeV range. There are a total of 290 sources in the Pulsar class, with 124 at high Galactic latitudes and 166 at low Galactic latitudes. They are labeled as as ``psr".
  \item [3.]
Non-AGN and non-pulsar class, which includes non-AGN galaxies (SBG, sbg, gal), supernova remnants and pulsar wind nebulae (PWN, pwn, SNR, snr, spp), star-forming regions (SFR, sfr), globular clusters (glc), binary systems (HMB, hmb, LMB, lmb, NOV, nov), novae (NOV), and galactic centers (GC). The non-AGN and non-pulsar sources are dominated by galactic sources and are mainly distributed in low Galactic latitude regions (209/264). They are labeled as ``other".
  \item [4.]
Unassociated class, which consists of 2157 unassociated sources and 134 weakly associated sources (unk). In these unassociated sources, 1166 are in low Galactic latitude regions and 1125 in high Galactic latitude regions.
\end{itemize}

Each source in the 4FGL-DR3 catalog contains 160 columns of data. After excluding descriptive columns, errors, missing values, historical data, duplicate data (i.e.  PLEC$\_$Index2), and inter-dependent parameters (e.g. GLAT, GLON vs  r.a., decl., and $\nu F_{\nu}$ vs  $F_{\nu}$), we obtained a total of 26 feature parameters directly for 4FGL-DR3 fits table (See Table \ref{Tab1} feature 1-26).  The parameters are mainly divided into four categories:
1)Positional features: These describe the celestial coordinates of the sources, including Galactic longitude and Galactic latitude.
2)Spectral features: These include spectral parameters derived from fitting the GeV data with \emph{PowerLaw},  \emph{LogParabola}, and \emph{PLSuperExpCutoff4}\footnote {See  \cite{2022ApJS..260...53A} for a detailed definition of three spectral types.}. They also encompass the significance differences when fitting with different spectral models.
3)Flux features: These involve the differential average flux and integrated flux in eight Fermi-LAT's energy bands\footnote {In 4FGL-DR3, the entire energy range of Fermi-LAT observations is divided into eight sub-bands: 0.05-0.1 GeV, 0.1-0.3 GeV, 0.3-1 GeV, 1-3 GeV, 3-10 GeV, 10-30 GeV,  and 30-1000 GeV.}, as well as the flux.
4) Significance features: These consist of the average significance and predicted photon event count of the sources.
5)Variability features: These include the variability index and fractional variability index.

To mitigate the systematic differences in flux and significance caused by the distances of sources, which could potentially mislead ML classification, we introduced some induced parameters as internal features.
Following \cite{2012ApJ...753...83A}, we provided the hardness ratios between eight Fermi-LAT bands to describe the soft and hard changes of spectrum, using
\begin{equation}\label{eq1}
 hr (i, j)=\frac {{\nu F_{\nu}}(j)-{\nu F_{\nu}}(i)}{{\nu F_{\nu}}(j)+{\nu F_{\nu}}(i)}
\end{equation}
where, $\nu F_{\nu} (i)$ and $\nu F_{\nu} (j)$ are the SEDs of different Fermi-LAT band, in which $\rm j = i + 1$. The quantity of $hr_{\rm ij}$ is always between -1 and 1.
Furthermore, to characterize the variation of the spectral index and the concavity of the spectrum, we define the concavity coefficient $H (i, j, k)$, as
\begin{equation}\label{eq2}
H (i, j, k)=hr (i, j) - hr (j, k)
\end{equation}
The quantity of $H_{\rm ijk}$ is always between -2 and 2.  So, there are seven hardness ratios and six concavity coefficients in a total of eight Fermi-LAT bands (See Table \ref{Tab1} feature 27-39).

In order to rationalize the dataset and optimize the machine learning workflow, it is important to pay attention to certain details: 
i)  To simplify the parameter space and reduce the computational load of ML, we took the logarithm of 18 features that spanned more than three orders of magnitude. The log10 flags of features can be found in Table \ref{Tab1}. 
ii) Due to its uniqueness, the source \emph{4FGL J1745.6-2859}, labeled as a galaxy center, was removed from the dataset. 
iii) The inverse Compton component of the Crab Nebula, denoted as \emph{4FGL J0534.5+2201i}, was removed from the dataset.  
iv) Less than 1$\%$ of the sources have parameter values of LP$\_$SigCurv, PLEC$\_$SigCurv, and $\nu F_{\nu} (8)$ equal to 0, which appear as infinitesimally small in logarithmic space. We have filled these values with the smallest non-zero value of the respective parameter.

Naturally, divided by a Galactic latitude threshold of $|b|=10^{\circ}$, the ML classification task is split into two frameworks. In the HGL region, the training set consists of 3407 AGNs, 124 pulsars, and 55 samples from the ``other" class. The objective of this classification is to identify a small number of non-AGN sources among 1125 unassociated sources, using active galactic nuclei as the background. In the LGL region, the training set comprises 407 active galactic nuclei, 166 pulsars, and 208 samples from the ``other" class. With a significantly larger number of unassociated sources (1166), a model is developed using a limited number of training samples for the purpose of three-class classification.

\section{Classification methods} \label{sec:classifier}

\subsection{ML Classifiers and Optimization Methods} \label{sec:classifier}

In the field of ML, there are many classification algorithms available. Examples include decision trees (DT), random forests (RF), logistic regression (LR), support vector machines (SVM), and multilayer perceptrons (MLP), among others, which are widely used for classifying unassociated sources in Fermi-LAT or evaluating the types of Fermi-LAT BCU (Blazar of unknown type). In this study, the chosen methods are LR, SVM, RF, and MLP.

LR is a commonly used statistical learning method \citep{cox1958regression}. It models the relationship between input features and class labels by establishing a logistic function (also known as the sigmoid function). It maps the feature space to the probability space, enabling classification. LR is relatively simple and has strong interpretability.

SVM is a classical ML algorithm that aims to construct an optimal hyperplane or maximize the margin in a multi-dimensional parameter space to achieve effective data classification \citep{cortes1995support}. It has the advantage of being able to handle high-dimensional data and non-linear problems while exhibiting good generalization capabilities.

DT is one of the earliest ML algorithms. It uses feature parameters to create nodes and makes branching decisions based on certain criteria \citep{breiman1984classification}. A large number of nodes form a tree-like structure. However, decision trees are prone to over-fitting when the depth increases. To address this issue, an early ensemble learning algorithm called random forest was developed. In random forest, multiple decision trees are combined using the bagging method \citep{breiman2001random}. The trees are trained on different subsets of the data, and the final result is determined by voting. This approach helps mitigate over-fitting and improves the model's generalization ability.

MLP is a simple artificial neural network model that consists of an input layer, hidden layers, and an output layer (e.g. \citealt{scikit-learn}). Each layer contains multiple neurons. The input layer receives raw data, the hidden layers are responsible for feature extraction and nonlinear transformations of the data, and the output layer produces the final prediction results. Each neuron has an activation function, commonly used activation functions include Sigmoid, ReLU, and Tanh. However, training and optimizing MLP can be complex, and it requires a substantial amount of data and computational resources for hyperparameter tuning and optimization.

Our dataset consists of a 39-dimensional parameter space. Having too many features can result in a large computational burden and potentially lead to a decrease in accuracy (e.g. \citealt{2019ApJ...872..189K}). To address this, we employ a model-dependent feature selection method called Recursive Feature Elimination (RFE), which allows us to select the optimal parameter space for different classification algorithms and scenarios. RFE works by training the model using all the features and iteratively removing the least important features based on the model's feedback. This process continues until an accuracy-feature count curve is obtained, from which the optimal feature subset can be determined.

All machine learning algorithms have model parameters that affect the training and performance of the model. However, there are certain parameters that cannot be learned during training and are referred to as hyper-parameters. For example,  hyper-parameters include the number of trees and maximum depth in random forests, the hidden layer structure and activation function in MLP. Hyper-parameters play a crucial role in model training and performance, but they need to be manually set and cannot be directly learned from the data. To determine the optimal hyper-parameter values, a common approach is grid search.

To train and optimize a model, it is necessary to partition the dataset into separate sets for training and testing the performance. However, the randomness of data partitioning, especially for imbalanced samples, can lead to unstable classifier performance. In order to ensure stable and reproducible classifier performance, we employed 5-fold stratified cross-validation in the training and optimization of all classifiers.

Due to the inconsistent kernels and classification principles of different classification algorithms, it is impossible for all algorithms to produce identical results for the same sample. To obtain a unified result, different methods can be employed: i) Seeking the union of predicted results: In this approach, inconsistent classification results are disregarded, and only the agreed-upon classifications are considered (e.g. \citealt{2021RAA....21...15Z, 2022A&A...660A..87B}). ii) Using a voting ensemble classifier: This method involves combining the predictions of multiple classifiers and determining the final classification based on a voting scheme. Through the votes of different classifiers, a consistent classification result is obtained. In this study, an ensemble voting classifier is employed to achieve a unified classification result. This approach takes advantage of the collective decision-making of multiple classifiers, which enhances the robustness and reliability of the final classification outcome.

The process of creating, training, optimizing, and testing all the classifiers mentioned above can be implemented using the Python library \emph{scikit-learn} \citep{scikit-learn}.

\subsection{Bayesian Gaussian model} \label{sec:classifier}

The Bayesian principle is a fundamental inference tool in the fields of statistics and ML (e.g. \citealt{gelman2013bayesian}). It is built upon the relationship between prior and posterior probabilities. The Bayesian principle plays a key role in various aspects such as parameter estimation, hypothesis testing, and model selection, providing a solid theoretical foundation for data analysis and inference.

In a multi-class classification problem, let there be classes $m_1$, $m_2$, ..., $m_n$, and given the multidimensional parameters $(x_1, x_2, ..., x_m)$. It is assumed that these parameters are mutually independent. Each parameter $(x_1, x_2, ..., x_m)$ in each class follows an independent Gaussian distribution $N(\mu_{i}, \sigma_{i}^2)$, where $\mu_{i}$ represents the mean and $\sigma_{i}$ represents the variance. The probability density function of the Gaussian distribution is given by:
\begin{equation}\label{eq3}
P(x_i|m_n) = \frac{1}{\sqrt{2\pi\sigma_{i}^2}} \cdot e^{-\frac{(x_i-\mu_{i})^2}{2\sigma_{i}^2}} \end{equation}
The prior probability $P(m_n)$ for each class can be estimated based on the sample counts:
\begin{equation}\label{eq4}
P(m_n) = \frac{N_{m_n}}{N_{\text{all}}}
\end{equation}
Here, $N_{m_n}$ represents the number of samples in class $m_n$, and $N_{\text{all}}$ represents the total number of samples. By fitting Gaussian distributions to all the parameter distributions and applying Bayes' theorem, the posterior probability of a sample belonging to class $m_n$ can be calculated as follows:
\begin{equation}\label{eq5}
P(m_n|(x_1,x_2,...,x_n)) = \frac{P(m_n)}{P(x_1,x_2,...,x_n)} \prod_{i=1}^{m} P(x_i|m_n) \end{equation}
The normalization factor $P(x_1,x_2,...,x_n)$ is defined as:
\begin{equation}\label{eq6}
P(x_1,x_2,...,x_n) = \sum_{j=1}^n (P(m_j) \prod_{i=1}^{m} P(x_i|m_j))
\end{equation}
For each sample, by substituting it into Equations \ref{eq3}-\ref{eq6}, the probabilities of it belonging to different classes, ($P_1$, $P_2$, ..., $P_n$), can be calculated. Here, $P_1+P_2+ \dots +P_n=1$. By comparing the relative values of these probabilities, the estimated class of the sample can be determined.

The following sections describes the the process of constructing classification models for the HGL and LGL regions.

\section{High Galactic Latitude region} \label{sec:high lati}

\begin{figure*}
\centering
 \includegraphics[height=9cm,width=5.8cm]{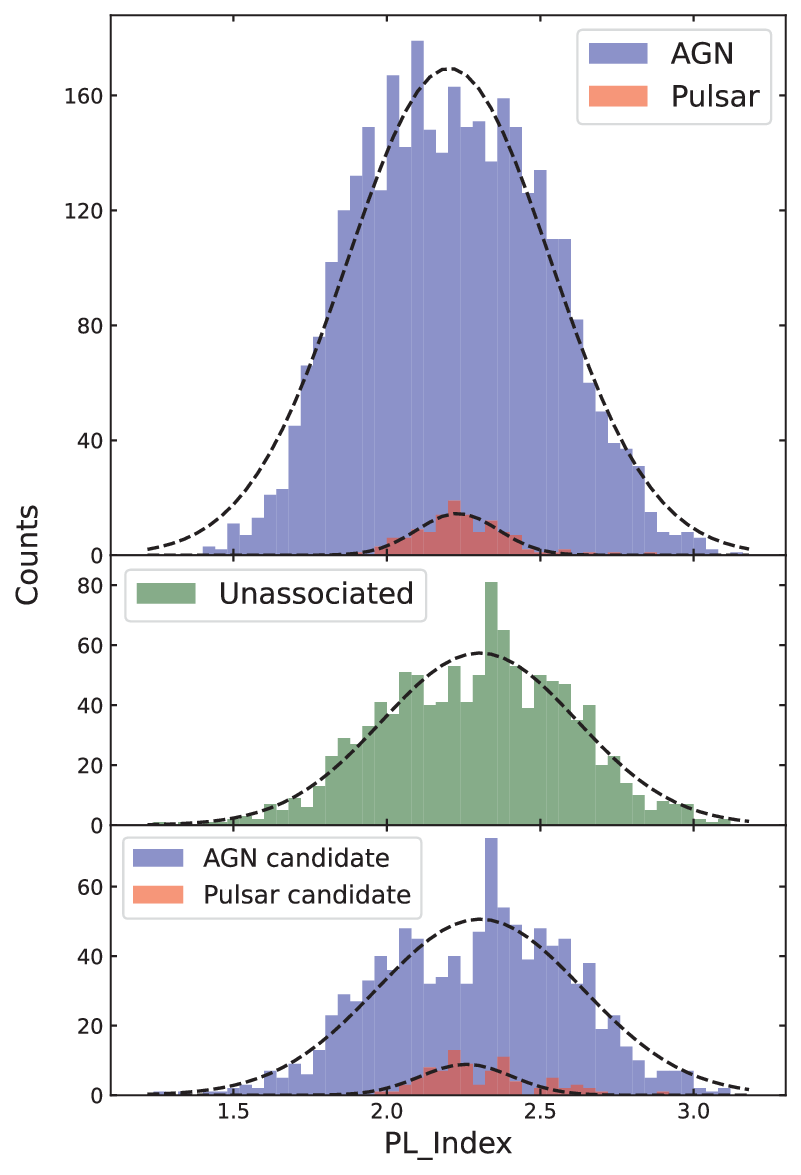}
  \includegraphics[height=9cm,width=5.8cm]{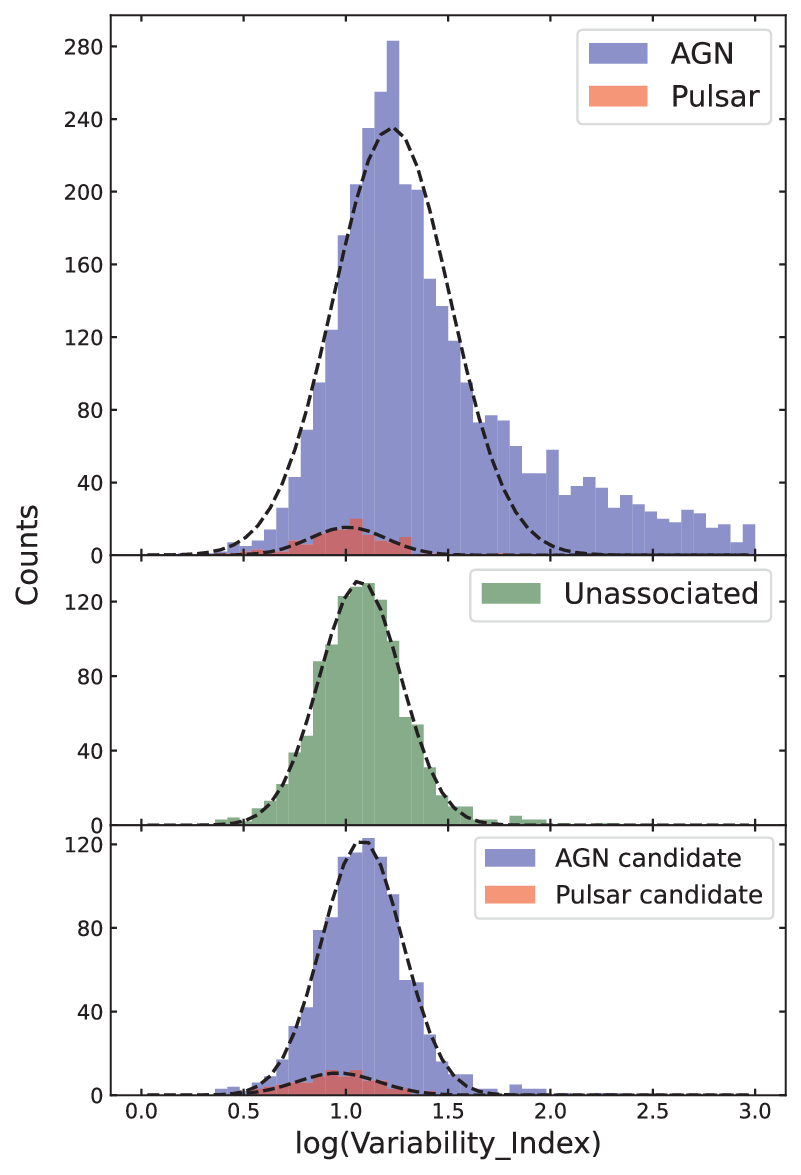}
    \includegraphics[height=9cm,width=5.8cm]{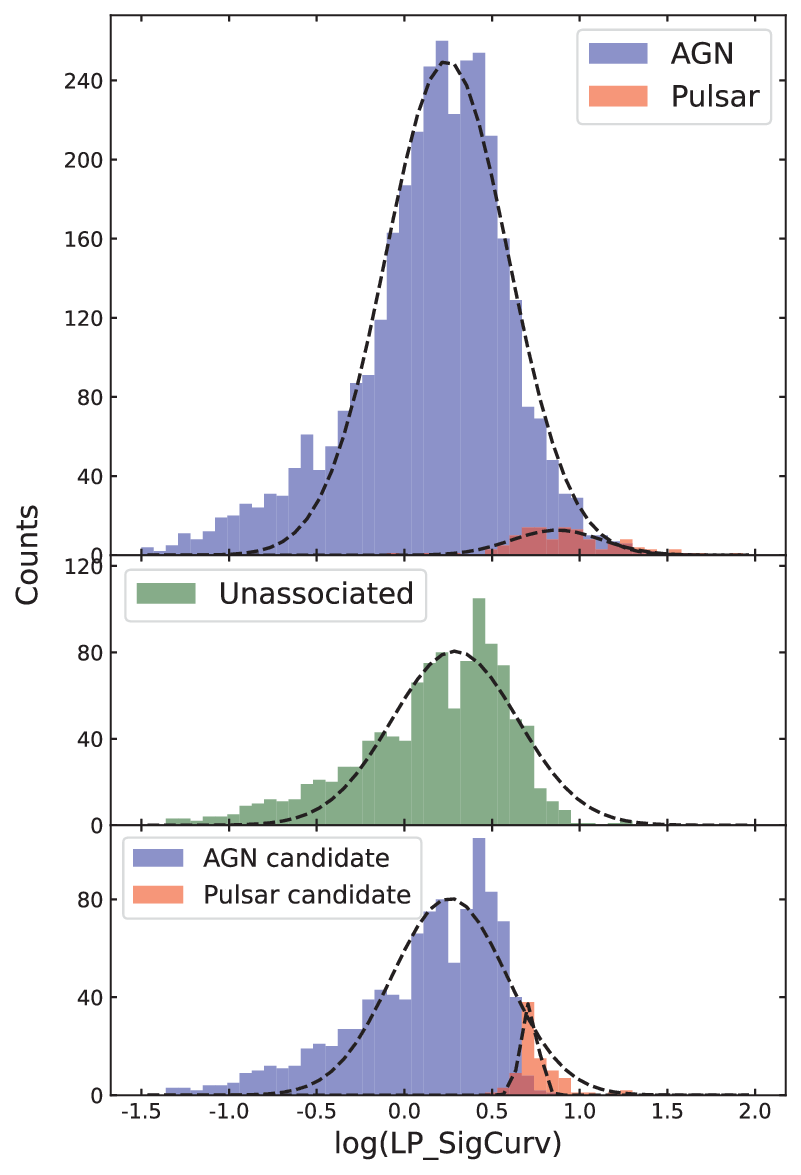}
\caption{Distribution plots of the gamma-ray spectral index, variability index, and log-parabolic fit significance for associated and unassociated sources in HGL. The top panel represents AGN-like and pulsar-like associated samples, the middle panel represents unassociated sources, and the bottom panel represents the results of the ML classification.}
 \label{fig1}
\end{figure*}

In the HGL dataset, there are 3407 AGN-like samples, 124 Pulsar-like samples, 55 other-like samples, and 1125 unassociated sources.  \cite{2022ApJS..260...53A} has discussed the distribution of gamma-ray spectral index in unassociated sources at high Galactic latitudes and finds that it resembles that of BCU sources, as well as all AGN-like objects (see Figure \ref{fig1} left panel), suggesting that most unassociated sources in the HGL region are likely AGNs. Here, we present the distributions of variability index and significances of log-parabolic fits for both associated and unassociated sources in the HGL region. As shown in Figure \ref{fig1} middle and right panel, the unassociated sources do not exhibit strong variability or significant spectral curvature compared to the associated sources. Their distribution is similar to that of AGNs, indicating that they are primarily AGN-likes. However, due to the overlap in some parameter distributions between AGNs and pulsars, it is not possible to rule out the possibility of non-AGN contamination.

Additionally, the number of other class samples is limited (55), and introducing them would significantly disrupt the distinguishing features of the samples and result in a significant decrease in classification accuracy. Therefore, a binary classification approach using AGN-like and pulsar-like categories is adopted for the HGL region.

\begin{figure*}
\begin{flushleft}
 \includegraphics[height=5.8cm,width=8.5cm]{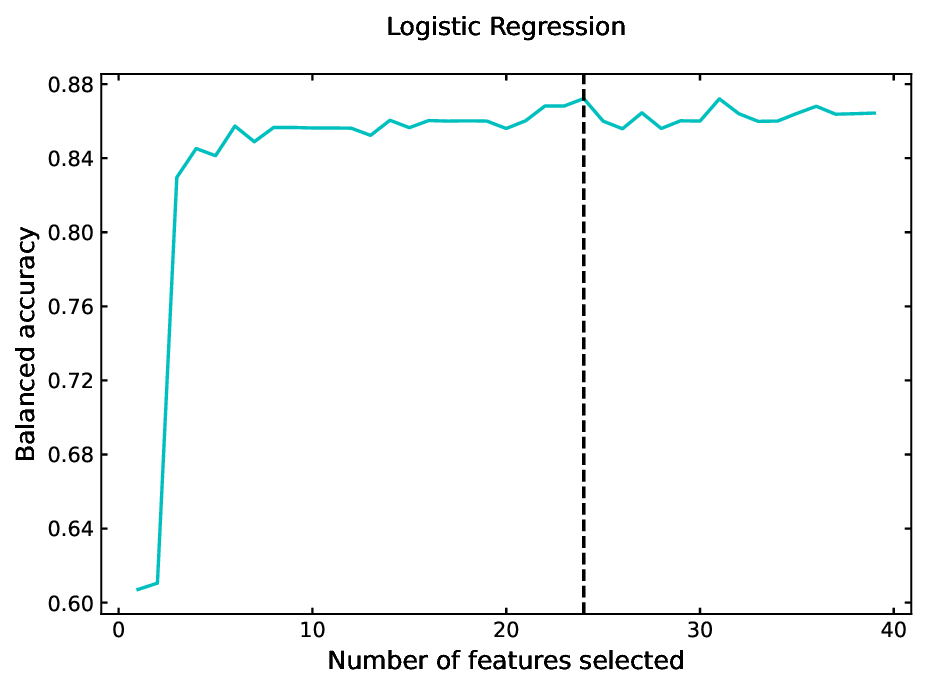}
  \includegraphics[height=5.8cm,width=8.5cm]{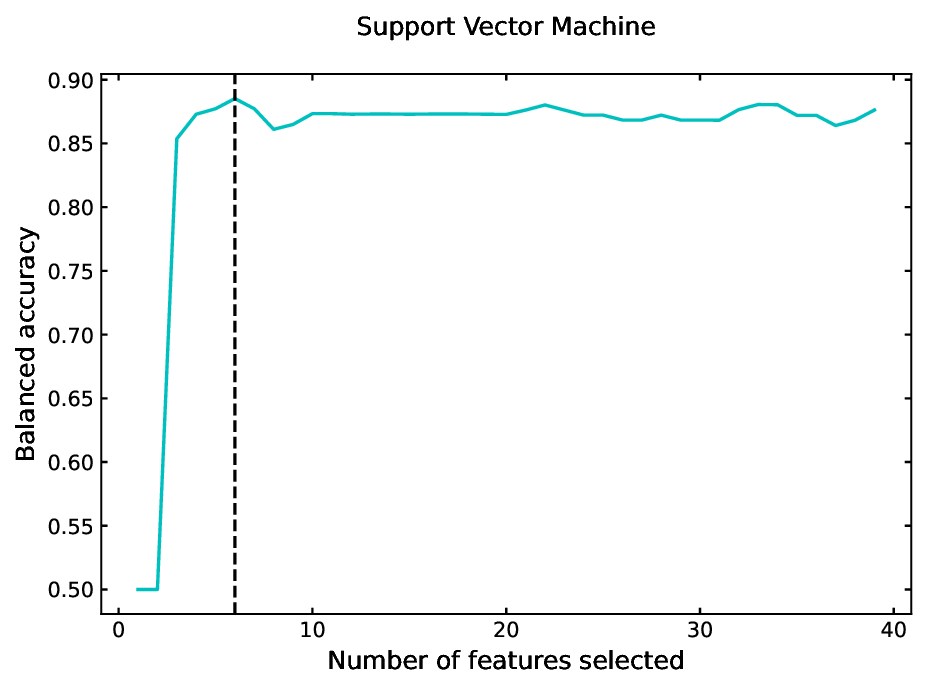}
    \includegraphics[height=5.8cm,width=8.5cm]{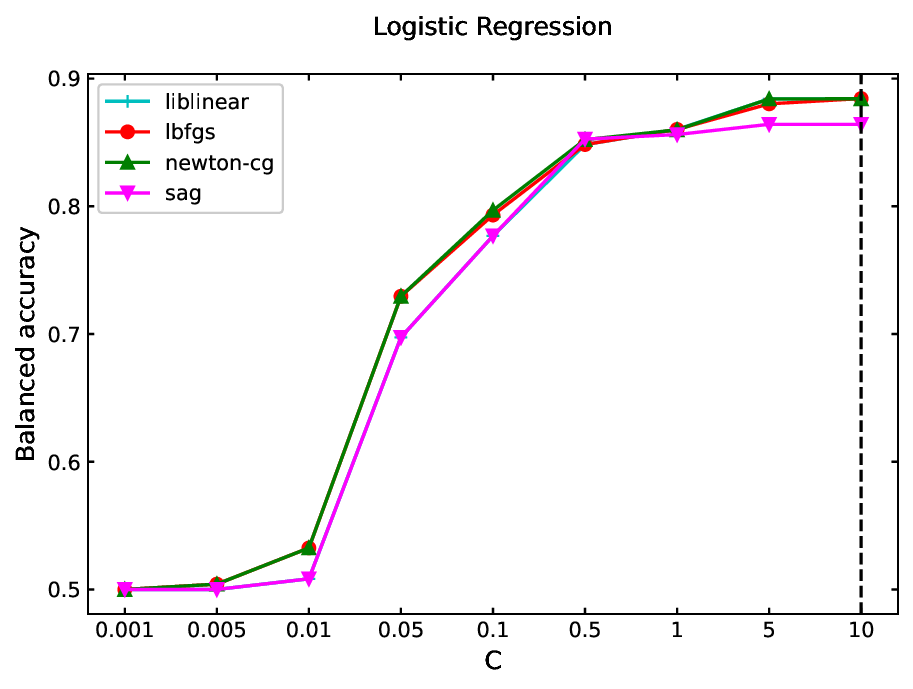}
      \includegraphics[height=5.8cm,width=8.5cm]{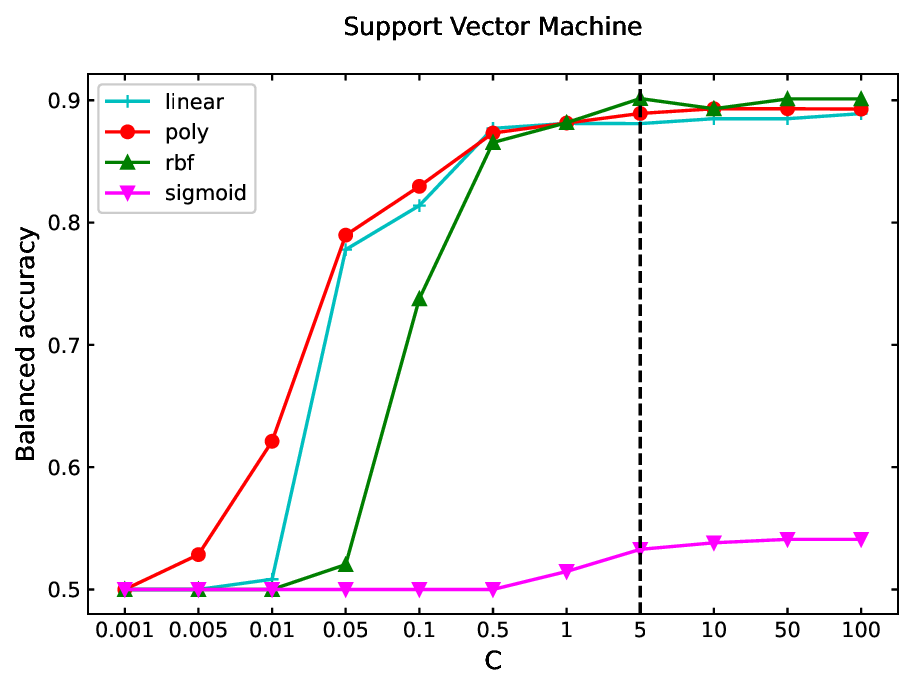}
       \includegraphics[height=5.8cm,width=8.5cm]{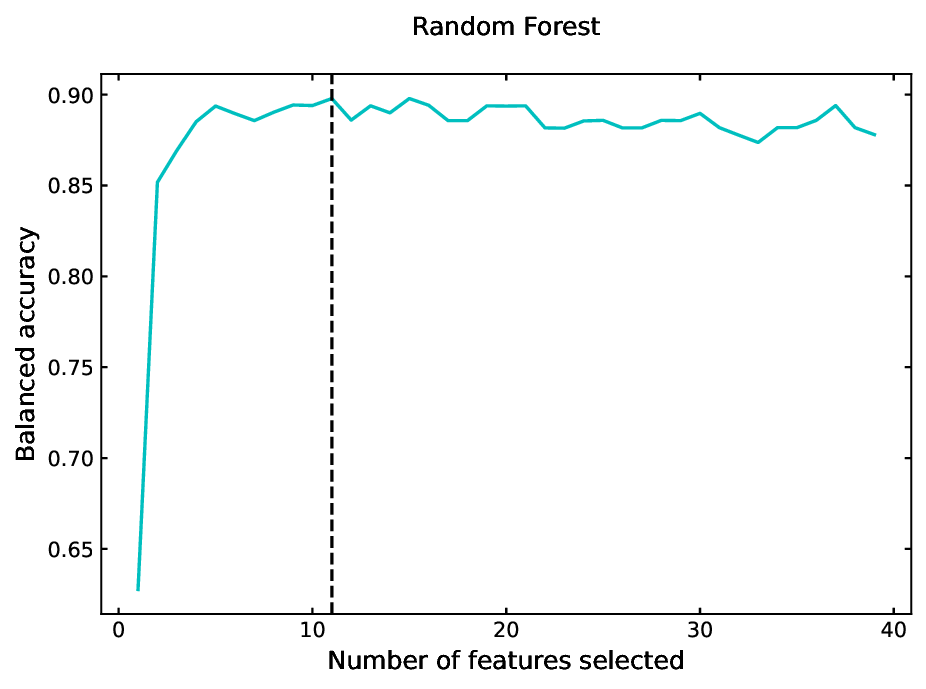}
        \includegraphics[height=5.8cm,width=8.5cm]{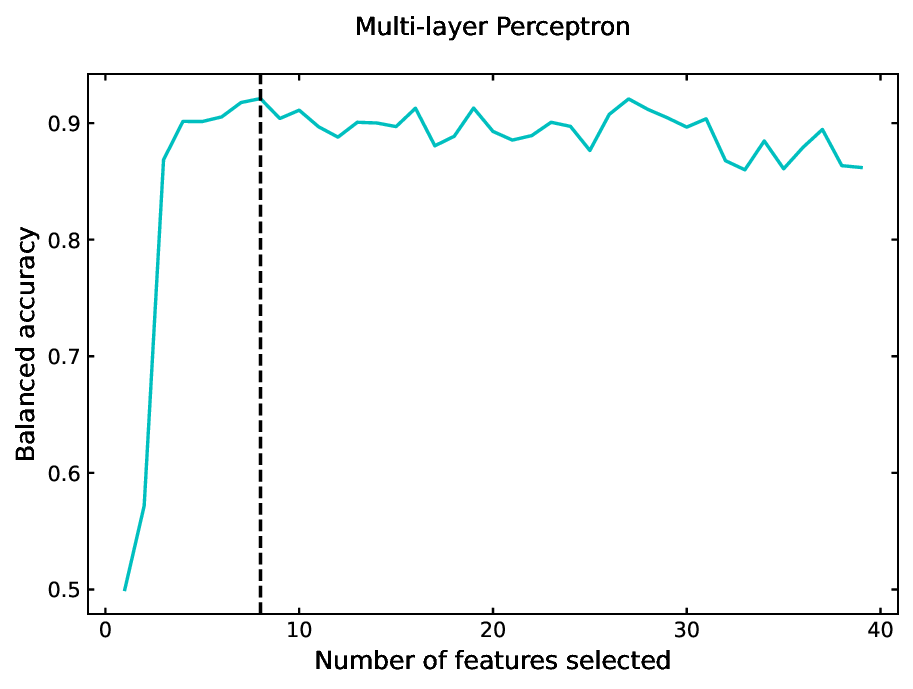}
            \includegraphics[height=5.8cm,width=8.5cm]{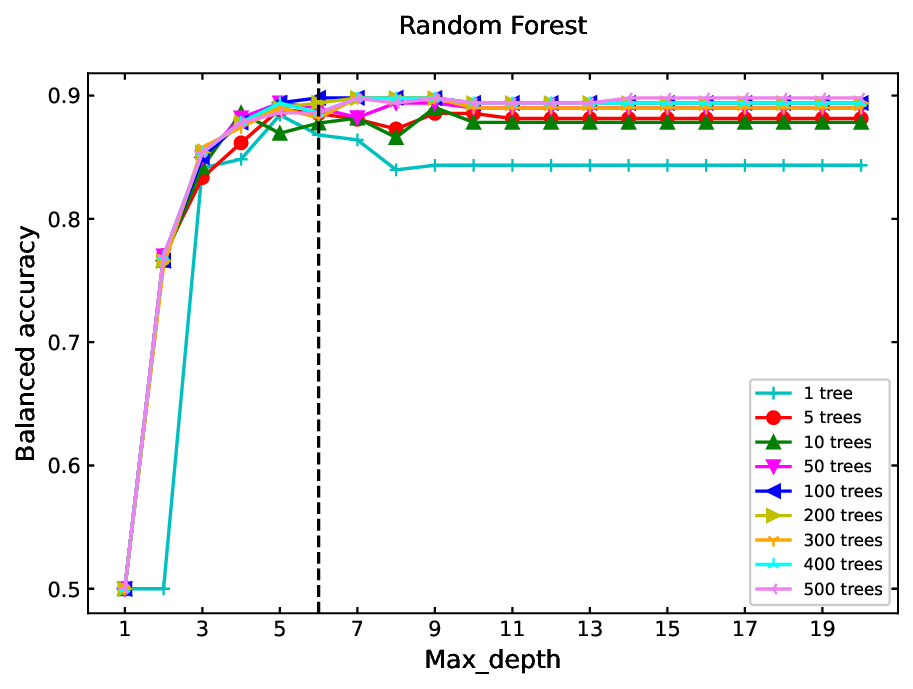}
\end{flushleft}
\caption{The RFE  curve plots and hyper-parameter gird search curve plots of four classifiers in HGL areas.}
 \label{fig2}
\end{figure*}

\subsection{Results of Individual ML model}

Due to the imbalanced nature of binary classification, we use balanced accuracy instead of simple  accuracy to evaluate the model. Balanced accuracy is an evaluation metric used for imbalanced classification problems. For n-class classification, its definition is as follows \citep{urbanowicz2015extra}:
\begin{equation}\label{eq7}
\rm{Balanced}\  \rm{accuracy}= \frac{1}{n}\sum_{i=1}^n ( \frac{\frac{\rm{TP}}{\rm{TP+FN}}+\frac{\rm{TN}}{\rm{TN+FP}}}{2})
\end{equation}
For each class, define it as the positive class, calculate the average of its sensitivity and specificity, and then calculate the average of all classes. When the samples of each class are perfectly balanced, the balanced accuracy is equal to the accuracy. The balanced accuracy for multiclass classification takes into account the imbalance among different classes, allowing for a more comprehensive evaluation of the model's performance in multi-class classification problems. In situations where there is sample distribution imbalance or varying importance among different classes, the balanced accuracy provides a more reasonable assessment of the model's performance.

Using the classifier optimization methods described in Section \ref{sec:classifier}, we trained and tested four different classification algorithms. The RFE curves for these four classifiers are shown in Figure \ref{fig2} top panel, and the corresponding optimal feature combinations are presented in Table \ref{Tab2}.

The grid search results for hyper-parameter tuning of LR, SVM, and RF are shown in Figure \ref{fig2} below panel \footnote{Please note that due to the large number of hyper-parameters in MLP, its grid search results cannot be represented in a 2-dimensional graph.}. From the figure, we can observe how classifier performance varies with different hyper-parameter values.  The optimal hyperparameter combinations for different algorithms are listed in Table \ref{Tab2}.

The classification results are shown in Table \ref{Tab2}. The four classification models exhibit consistent results, with a balanced accuracy of approximately 90$\%$. Among them, LR has a slightly lower balanced accuracy, while MLP has a slightly higher balanced accuracy. The classification yields approximately 1032 - 1059 samples classified as AGN-like and 66 - 93 samples classified as pulsar-like.

\begin{table*}
\centering
\caption{The information of the high-latitude classification models}\label{Tab2}
\resizebox{\textwidth}{!}{
\begin{tabular}{cccccc}
\hline \hline
\multirow{2}{*}{Estimator}&\multirow{2}{*}{Input features}& \multirow{2}{*}{Hyper-parameter}&\multirow{2}{*}{Test balanced accuracy}&\multirow{2}{*}{AGN}&\multirow{2}{*}{PSR}	\\
&&&\\
\normalsize(1) & \normalsize(2) & \normalsize(3) &\normalsize(4)  &\normalsize(5)&\normalsize(6)\\
\hline
\multicolumn{6}{c}{Individual classification estimator}\\	
\hline
\multirow{2}{*}{Logistic Regression}&3, 4, 6, 7, 8, 9, 10, 12, 13, 14, 15, 16, 17,	&		$C:10$&\multirow{2}{*}{$0.884\pm 0.027$}&\multirow{2}{*}{1042}&\multirow{2}{*}{83}\\
& 18, 21, 22, 23, 30, 31, 32, 35, 36, 37, 38 &$solver: `lbfgs$'&\\
\multirow{2}{*}{Support Vector Machine}&	\multirow{2}{*}{3, 8, 13, 15, 31} 	&	$C:5$	&\multirow{2}{*}{$0.901\pm 0.034$}&\multirow{2}{*}{1055}&\multirow{2}{*}{70}\\
&&$kernel: `rbf$'&\\
\multirow{2}{*}{Random Forest}&	\multirow{2}{*}{6, 8, 10, 12, 13, 14, 17, 23, 31, 36, 37}	&	$max\_depth: 6$	&\multirow{2}{*}{$0.898\pm 0.026$}&\multirow{2}{*}{1059}&\multirow{2}{*}{66}\\
&&$n\_estimators: 100$&\\
\multirow{4}{*}{Multilayer Perceptron}&	\multirow{4}{*}{8, 14, 28, 29, 30, 32, 36, 37}&	$max\_iter:500$	&\multirow{4}{*}{$0.926\pm 0.028$}&\multirow{4}{*}{1032}&\multirow{4}{*}{93}\\
& &$activation: `tanh$'&\\
&&$hidden\_layer\_sizes: (60, 60, 10)$&\\
&&$solver: `adam$'&\\
\hline
\multicolumn{6}{c}{Ensemble classification estimator}\\	
\hline
\multirow{2}{*}{Voting ensemble classifier}&	\multirow{2}{*}{}	&	$voting: `soft$' 	&\multirow{2}{*}{$0.918\pm 0.029$}&\multirow{2}{*}{1037}&\multirow{2}{*}{88}\\
&&$weights:[1, 2, 1, 4]$&\\
\hline
\end{tabular}}\\
{\footnotesize{{Note. Column (1): classifier Name; Column (2): optimal input features obtained through RFE; Column (3): best combination of hyper-parameters obtained from gird search; Column (4): test balanced accuracy from five-fold stratified cross-validation; Columns (5)-(6): Number of candidate objects obtained.  }}}
\end{table*}

\subsection{Results of Ensemble ML model}

By using an ensemble voting classifier, we combined the results of four individual classifiers. We performed a grid search on the weights of the four sub-classifiers in a ``soft" ensemble voting classifier. Through five-fold stratified cross-validation, we identified the optimal weights for the ensemble voting classifier. Among multiple optimal weight combinations, we selected the combination with the simplest weight sum (i.e., the simplest model).

With the optimal weights [1, 2, 1, 4] as hyper-parameters, the balanced accuracy reached $0.918 \pm 0.029$ in cross-validation (See Table \ref{Tab2}). We used an ensemble voting classifier to evaluate the categories of 1125 unassociated sources, resulting in 1036 AGN-like candidates and 89 pulsar-like candidates. Probability assessments were also conducted for the unassociated sources, leading to the creation of a probability catalog.

To assess the validity of the classification results, we examined the parameter distributions of the candidate objects, as shown in Figure \ref{fig1} bottom panel. The distribution of the gamma-ray spectral index, variability index and significance of log-parabolic fits for both types of candidates are not significant different with associated ones. Specifically, the pulsar-like candidates exhibited weaker variability compared to the AGN-like candidates, while they showed a higher degree of spectral curvature. This consistency with the parameter distributions of known AGNs and pulsars confirmed the reliability of classification results.

\section{Low Galactic Latitude region} \label{sec:low lati}

\begin{figure*}
\centering
 \includegraphics[height=8cm,width=5.8cm]{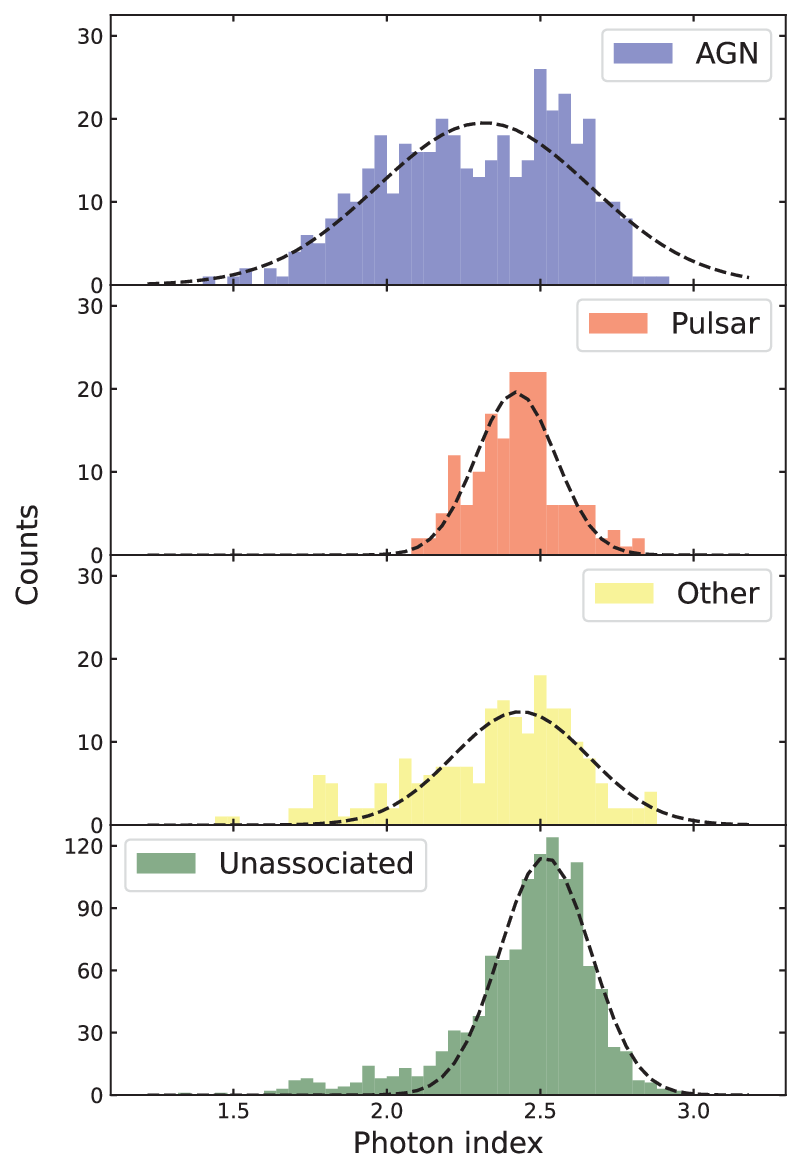}
  \includegraphics[height=8cm,width=5.8cm]{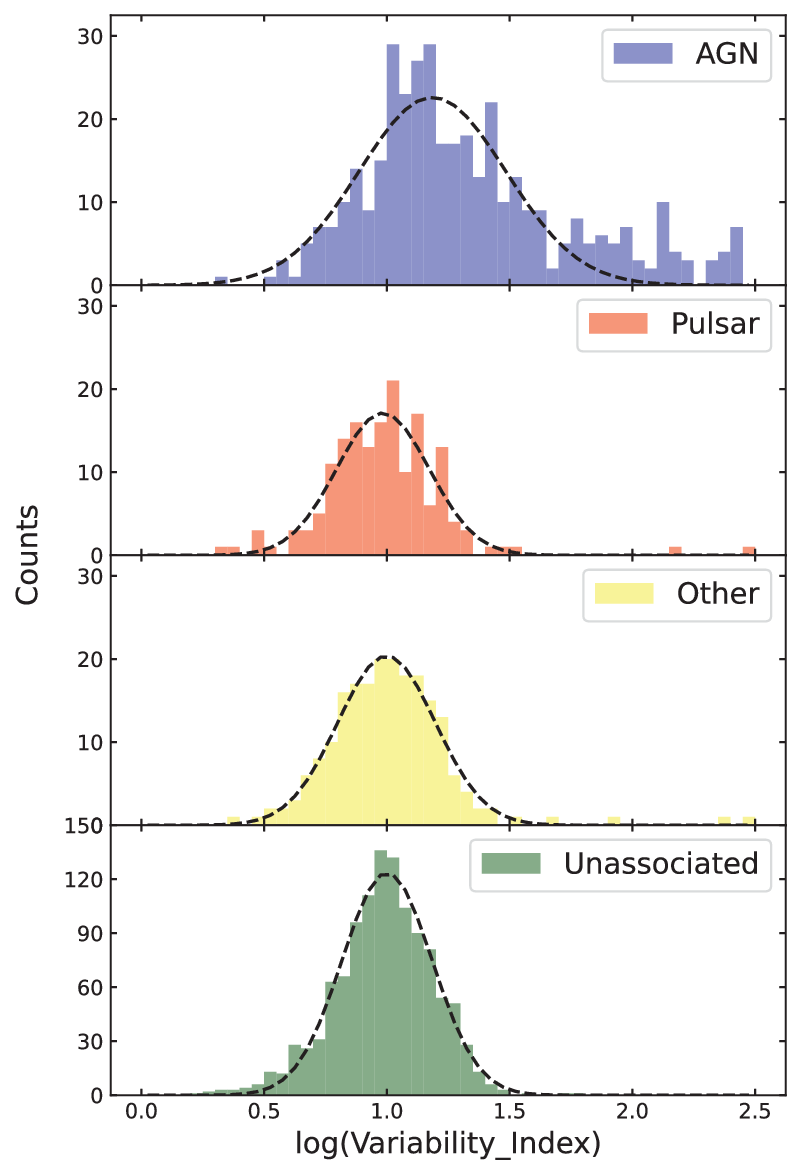}
    \includegraphics[height=8cm,width=5.8cm]{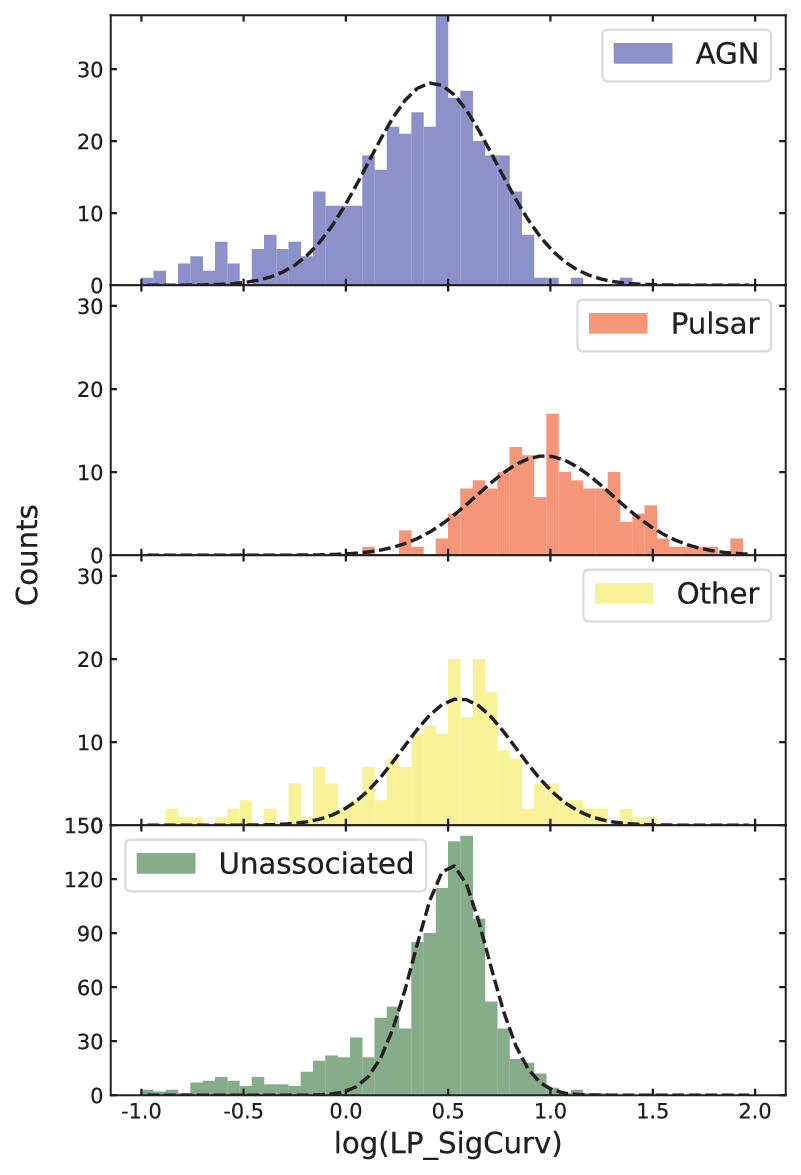}
\caption{Distribution plots of the gamma-ray spectral index, variability index, and log-parabolic fit significance for associated and unassociated sources in LGL. }
 \label{fig3}
\end{figure*}

In the LGL dataset, there are 407 AGN-like, 166 pulsar-like, 208 samples from the ``other" class, and 1166 unassociated sources. Due to the strong diffuse gamma-ray background radiation near the Galactic plane, the features of sources are not clear. Moreover, the number of unassociated sources is larger than that of associated sources, making it challenging to build a high-performance classifier.

We provide the distributions of gamma-ray photon index, variability index, and significances of log-parabolic fits for both associated and unassociated sources in the LGL dataset. From Figure \ref{fig3}, it can be seen that the unassociated sources in the LGL dataset are predominantly characterized by soft spectrum, weak variability, and moderate to weak significances of spectral curvature. Figure \ref{fig3} showed that the parameter distribution of unassociated sources in the LGL dataset partially overlaps with the AGN-like and Pulsar-like classes, and is most similar to the other-like class. The results indicated that the unassociated sources exhibit less significant spectral curvature, unlike pulsars, and their variability is extremely weak, unlike AGNs. These results suggested that unassociated sources in the LGL dataset are likely dominated by the other-like class rather than the pulsar and AGN classes.

Comparing the gamma-ray spectral index distribution of AGNs near the Galactic plane with those at HGL, it is evident that there is an excess of soft spectral samples (with $\Gamma > 2.4$) in the LGL region, as shown in Figure \ref{fig3} (left panel). According to \cite{2022ApJS..260...53A}, the estimated number of these excess blazars is $75 \pm 4$, which could be attributed to contamination from Galactic components.

Based on the detected counts of blazars at high latitudes and accounting for the detected flux difference due to the brighter diffuse emission background near the Galactic plane, the number of blazars with $|b| < 10^\circ$ is estimated to be $340 \pm 20$  \citep{2022ApJS..260...53A}. Considering the 1037 AGN-like candidates provided by the HGL ML analysis, we can estimate the number of AGNs in the low-latitude region using the following equation:
\begin{equation}\label{eq8}
N_{\rm{agn}}^{e}=N_{\rm{bla}}^{e}\frac{N_{\rm{bla}}+N_{\rm{nonbla}}+N_{\rm{agn}}^{\rm{can}}}{N_{\rm{bla}}}
\end{equation}
Here, $N_{\rm{bla}}=3342$ represents the current number of blazars in the HGL region, $N_{\rm{bla}}+N_{\rm{nonbla}}=3407$ is the total number of AGNs at HGL region, $N_{\rm{agn}}^{\rm{can}}=1036$ denotes the number of AGN candidates in the high-latitude region, and $N_{\rm{bla}}^{e}=340 \pm 20$ represents the estimated number of blazars in the low-latitude region based on existing observations \citep{2022ApJS..260...53A}. According to our estimation, there are approximately $452 \pm 27$ observable AGNs in the LGL region. The sum of the obtained AGN-like candidates and the existing AGNs should roughly satisfy this constraint.

Due to the near saturation of associated AGN counts at low Galactic latitudes and the presence of an excess of soft-spectrum sources leading to sample impurity, we first employ a Bayesian-Gaussian model to screen the training samples.

\subsection{Bayesian Gaussian estimation}

In \cite{2022ApJS..260...53A}, the variability index and the significance of the log-parabolic fit were employed to differentiate between AGNs and pulsars. This study focused specifically on the excess of soft-spectrum index sources in LGL AGN. By utilizing these three parameters, a BG classifier is established for probabilistic inference in AGN-nonAGN classification.

First, the known samples from the LGL training dataset are divided into two categories: AGN and non-AGN. Then, Gaussian function fitting is performed separately for the distributions of the three parameters (gamma-ray spectral index, variability index, and logarithm of the parabolic fit) for both AGN and non-AGN samples. These distributions are represented as $N(\mu_i, \sigma_i^2)$. For a given sample to be classified, its parameter values and the obtained parameter distributions $N(\mu_i, \sigma_i^2)$ are substituted into Equations \ref{eq3}-\ref{eq6}, which calculate the likelihood probabilities of it belonging to the AGN or non-AGN category. These probabilities are denoted as $L_a$ and $L_{na}$, respectively. It should be noted that $L_{a} + L_{na} = 1$. By comparing the relative magnitudes of $L_a$ and $L_{na}$, the closeness of the sample to AGN or non-AGN in the parameter space can be assessed. This method is initially used to estimate the reliability of associated AGN samples in the LGL region.

\begin{figure}
\centering
 \includegraphics[height=12cm,width=8.2cm]{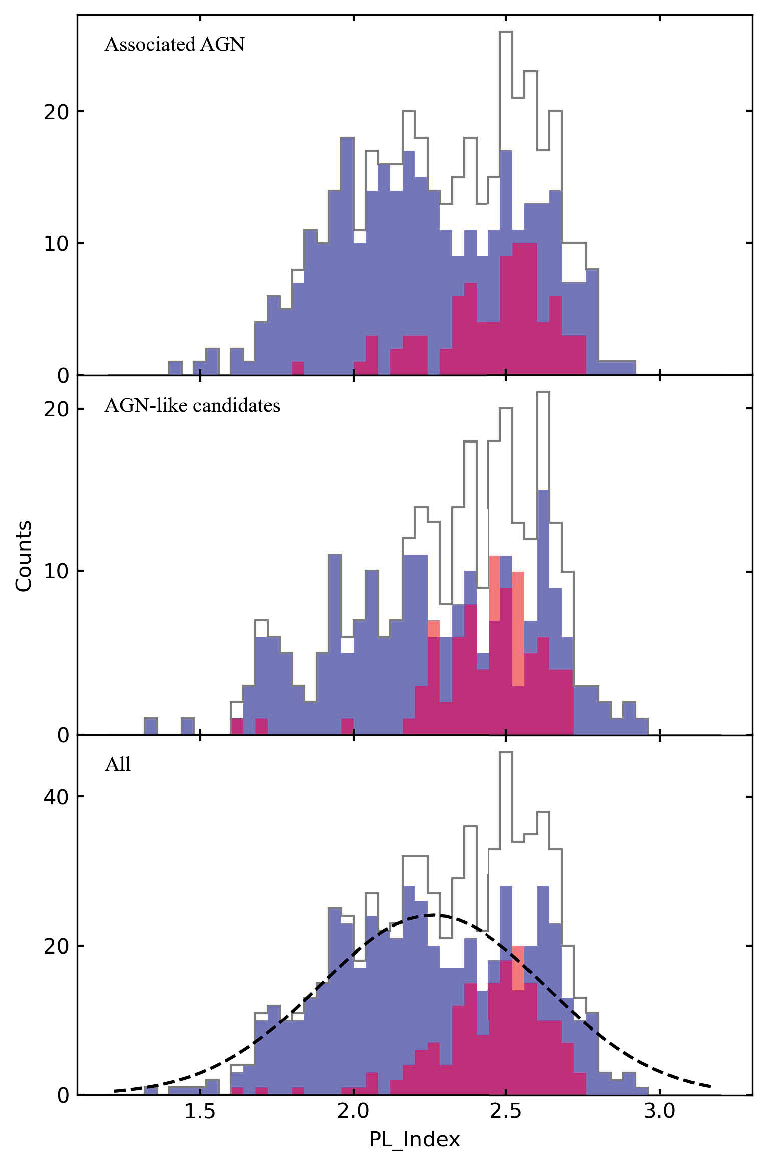}
\caption{The spectral index distribution of the Bayesian Gaussian parameter evaluation results for low-latitude AGN-like candidates and associated AGNs. The gray line represents all sources, the red region represents sources with low confidence according to Bayesian Gaussian parameter evaluation, and the blue region represents the remaining samples.}
 \label{fig4}
\end{figure}

When applying the BG model to the associated LGL AGN, it was found that 81 sources, characterized by excessively soft spectra and weak variability (the red region in Figure \ref{fig4} middle panel), were considered dissimilar to AGN with a classification threshold of $L_{na} >0.5>L_{a}$. This result was highly consistent with the $75\pm 4$ sources reported in 4FGL-DR3 \citep{2022ApJS..260...53A}. These sources were likely to be misassociations in the Fermi-LAT catalog. The information for these 81 sources was provided in a machine-readable format for further analysis.

After excluding these low-confidence samples, there remain a total of 326 AGNs in the low-latitude region. The complete distribution of their spectral indices can be observed in the bottom panel of Figure \ref{fig4}, where an excess of soft-spectrum AGNs has been suppressed. These samples, combined with 166 pulsars and 208 other-like sources, were used as the training set to construct a ternary ML classifier for the low-latitude region.

\begin{figure*}
\begin{flushleft}
 \includegraphics[height=5.8cm,width=8.5cm]{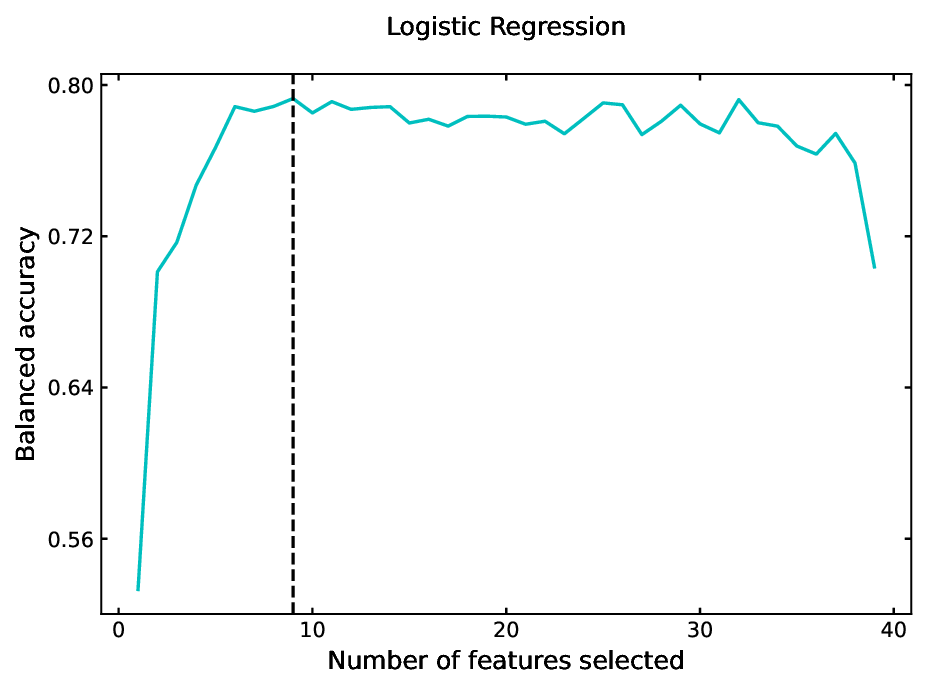}
  \includegraphics[height=5.8cm,width=8.5cm]{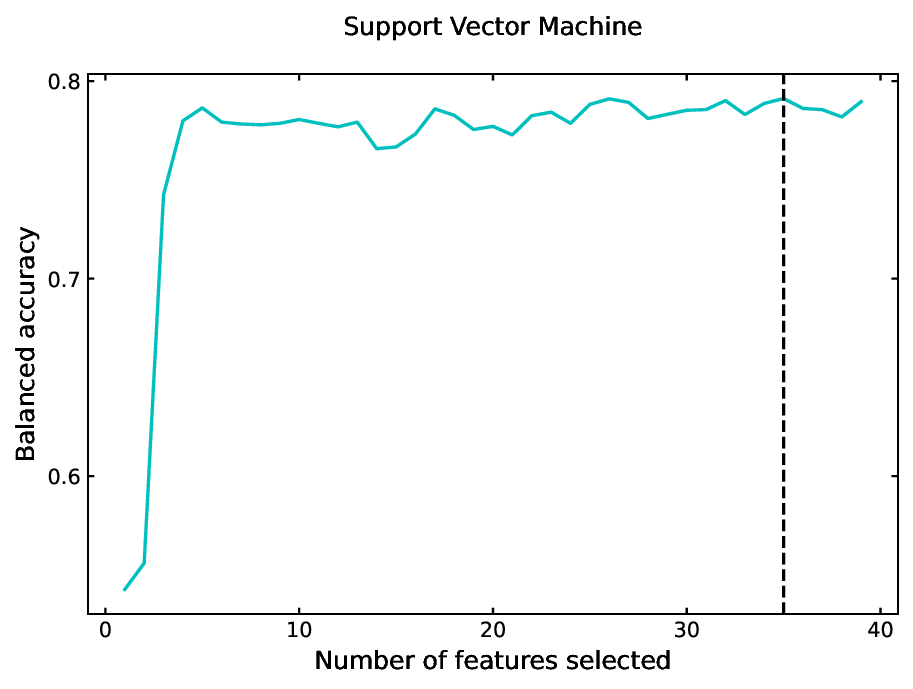}
    \includegraphics[height=5.8cm,width=8.5cm]{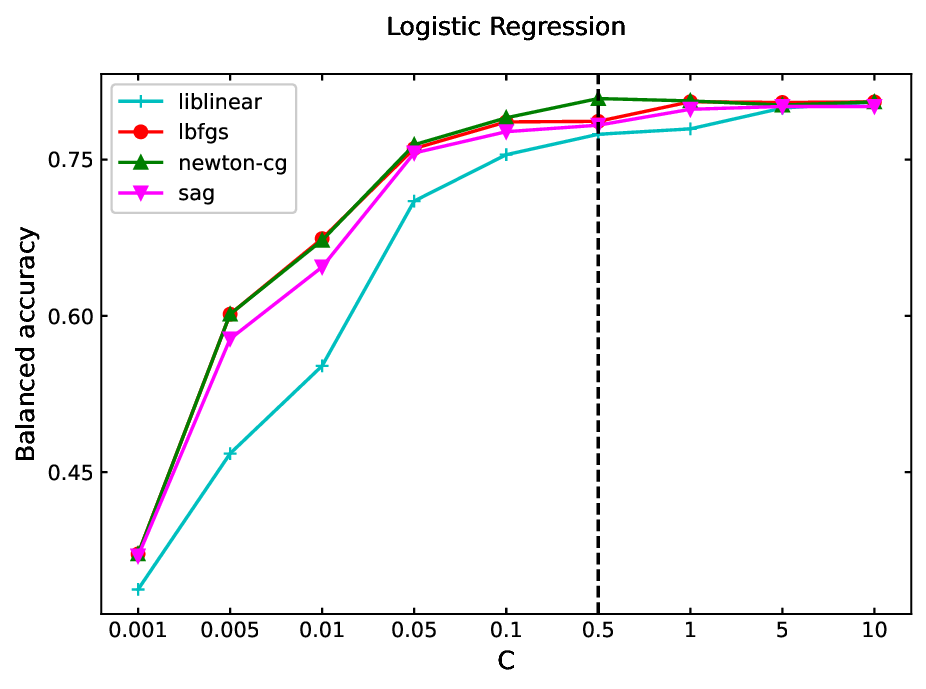}
      \includegraphics[height=5.8cm,width=8.5cm]{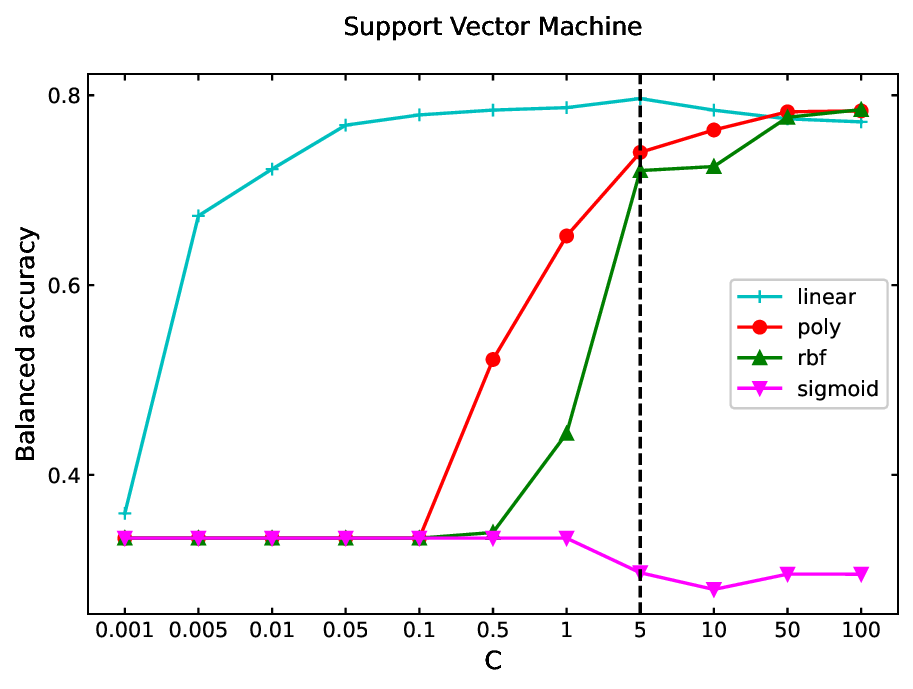}
       \includegraphics[height=5.8cm,width=8.5cm]{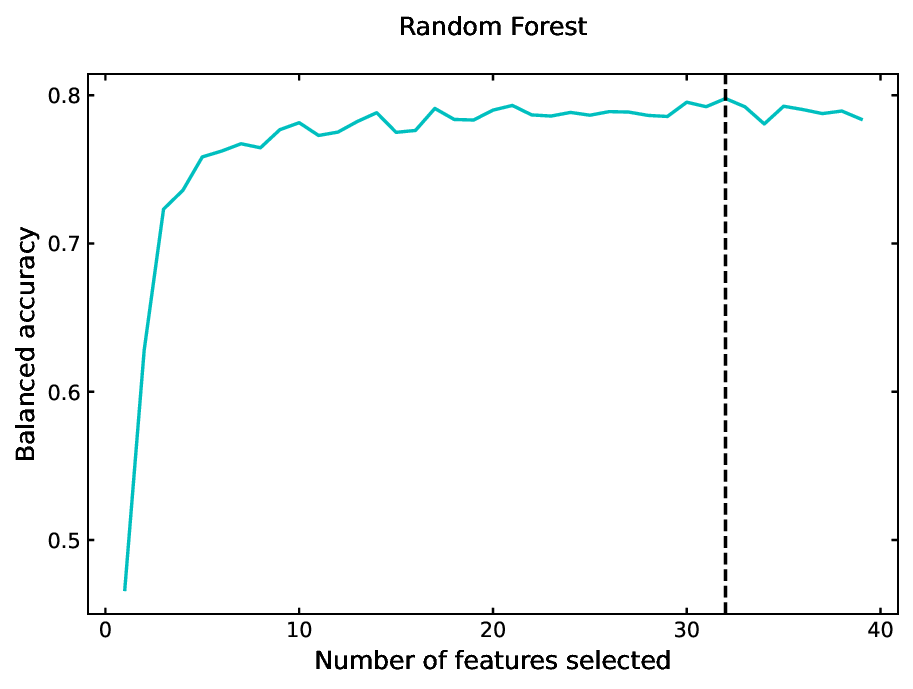}
         \includegraphics[height=5.8cm,width=8.5cm]{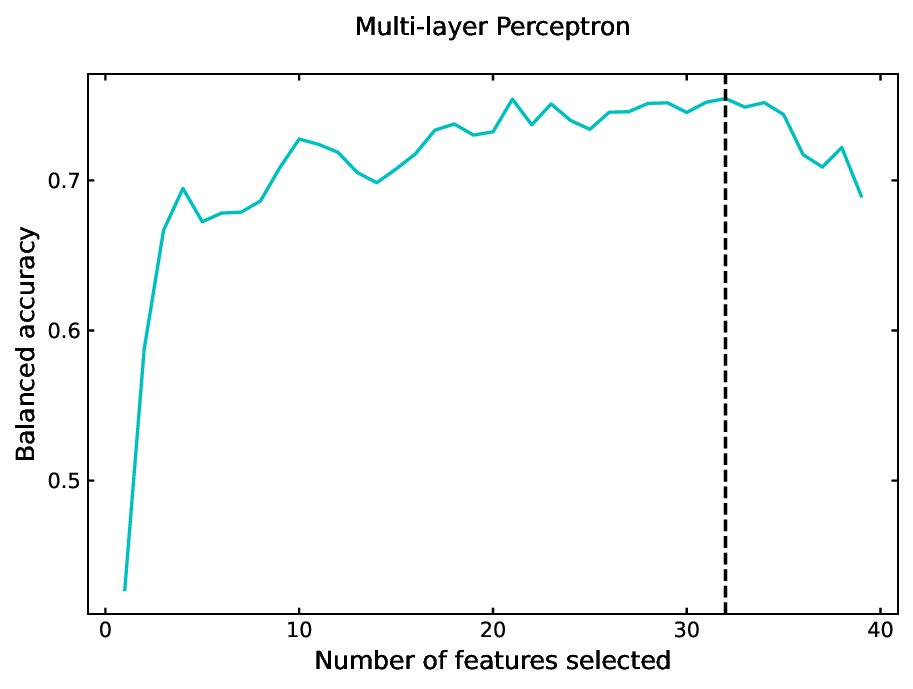}
            \includegraphics[height=5.8cm,width=8.5cm]{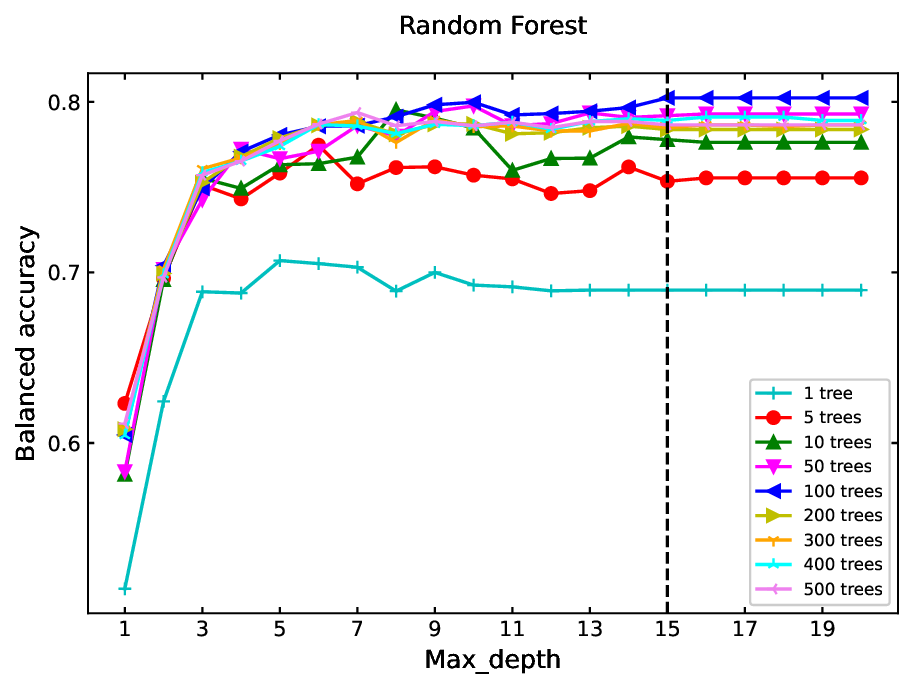}
\end{flushleft}
\caption{The RFE  curve plots and hyper-parameter gird search curve plots of four classifiers in LGL areas.}
 \label{fig5}
\end{figure*}

\subsection{Supervised machine learning classification}

\subsubsection{Results of Individual ML model}

\begin{table*}
\centering
\caption{The information of the low-latitude classification models}\label{Tab3}
\resizebox{\textwidth}{!}{
\begin{tabular}{ccccccc}
\hline \hline
\multirow{2}{*}{Estimator}&\multirow{2}{*}{Input features}& \multirow{2}{*}{Hyper-parameter}&\multirow{2}{*}{Test balanced accuracy}&\multirow{2}{*}{AGN}&\multirow{2}{*}{PSR}	&\multirow{2}{*}{OTHER}\\
&&&\\
\normalsize(1) & \normalsize(2) & \normalsize(3) &\normalsize(4)  &\normalsize(5)&\normalsize(6)&\normalsize(7)\\
\hline
\multicolumn{6}{c}{Individual classification estimator}\\	
\hline
\multirow{2}{*}{Logistic Regression}&	\multirow{2}{*}{3, 8, 13, 14, 15, 16, 17, 22}	&	$C:0.5$	&\multirow{2}{*}{$0.809\pm 0.021$}&\multirow{2}{*}{318}&\multirow{2}{*}{136}&\multirow{2}{*}{712}\\
& &$solver: `newton-cg$'&\\
\multirow{2}{*}{Support Vector Machine}&	\multirow{2}{*}{All except 1, 2, 25, 26} 	&$C:5$		&\multirow{2}{*}{$0.797\pm 0.039$}&\multirow{2}{*}{306}&\multirow{2}{*}{155}&\multirow{2}{*}{706}\\
&&$kernel: `linear$'&\\
\multirow{2}{*}{Random Forest}&	\multirow{2}{*}{All except 19, 32, 33, 34, 35, 38, 39} 	&$max\_depth: 15$	&\multirow{2}{*}{$0.802\pm 0.021$}&\multirow{2}{*}{300}&\multirow{2}{*}{106}&\multirow{2}{*}{760}\\
&&$n\_estimators: 100$&\\
\multirow{4}{*}{Multilayer Perceptron}&	\multirow{4}{*}{All except 1, 21, 22, 23, 24, 26, 27}	&	$max\_iter:500$	&\multirow{4}{*}{$0.788\pm 0.046$}&\multirow{4}{*}{260}&\multirow{4}{*}{199}&\multirow{4}{*}{707}\\
&&$activation: `relu$'&\\
&&$hidden\_layer\_sizes: 100, 30, 50$&\\
&&$solver: `adam$'&\\
\hline
\multicolumn{6}{c}{Ensemble classification estimator}\\	
\hline
\multirow{2}{*}{Voting ensemble classifier}&	\multirow{2}{*}{}	&	$voting: `soft$' 	&\multirow{2}{*}{$0.815\pm 0.027$}&\multirow{2}{*}{290}&\multirow{2}{*}{135}&\multirow{2}{*}{741}\\
&&$weights:[4,1,1,1]$&\\
\hline
\end{tabular}}\\
{\footnotesize{{Note. Column (1): classifier Name; Column (2): optimal input features obtained through RFE; Column (3): best combination of hyper-parameters obtained from gird search; Column (4): test balanced accuracy from five-fold stratified cross-validation; Columns (5)-(7): Number of candidate objects obtained. }}}
\end{table*}

Using the classifier optimization methods described in Section 3, we trained and tested four different classification algorithms.

The RFE curves for these four classifiers are shown in Figure \ref{fig5} top panel, and the corresponding optimal feature combinations are presented in Table \ref{Tab3}. The grid search results for hyper-parameter tuning of LR, SVM, and RF are shown in Figure \ref{fig5} bottom panel. From the figure, we can observe how classifier performance varies with different hyper-parameter values.  The optimal hyperparameter combinations for different algorithms are listed in Table \ref{Tab3}.

The classification results are shown in Table \ref{Tab3}. The results from the four classification models are consistent, with a balanced accuracy of approximately 80$\%$. Among the four classifiers, LR and RF achieved slightly higher accuracy, while MLP had lower accuracy. The classification results indicate that in the low-latitude region, there are approximately 260-318 samples classified as AGN-like, 106-199 samples classified as pulsar-like, and 706-760 samples belonging to other-like classes. The evaluations of sources varied greatly among different algorithms, highlighting the lack of reliability in the results obtained from a single classifier.

To obtain a unified result, we used a voting ensemble classifier to combine the results of multiple classifiers.

\subsubsection{Results of Ensemble ML model}

Using an ensemble voting classifier, we combined the results of four individual classifiers. We performed a grid search to optimize the weights of the four sub-classifiers in the "soft" ensemble voting classifier. Through five-fold stratified cross-validation, we determined the optimal weights for the ensemble voting classifier.

Using the best weights [4,1,1,1] as hyper-parameters, the balanced accuracy reached $0.815 \pm 0.027$ in cross-validation (see Table \ref{Tab3}). We evaluated the ensemble voting classifier on 1166 unassociated sources, resulting in 290 AGN-like candidate sources, 135 pulsar-like candidate sources, and 741 candidates from other categories. From the weights of the ensemble classifier, it can be known that the LR classifier dominates in the voting process.

\begin{figure}
\centering
 \includegraphics[height=12cm,width=8.2cm]{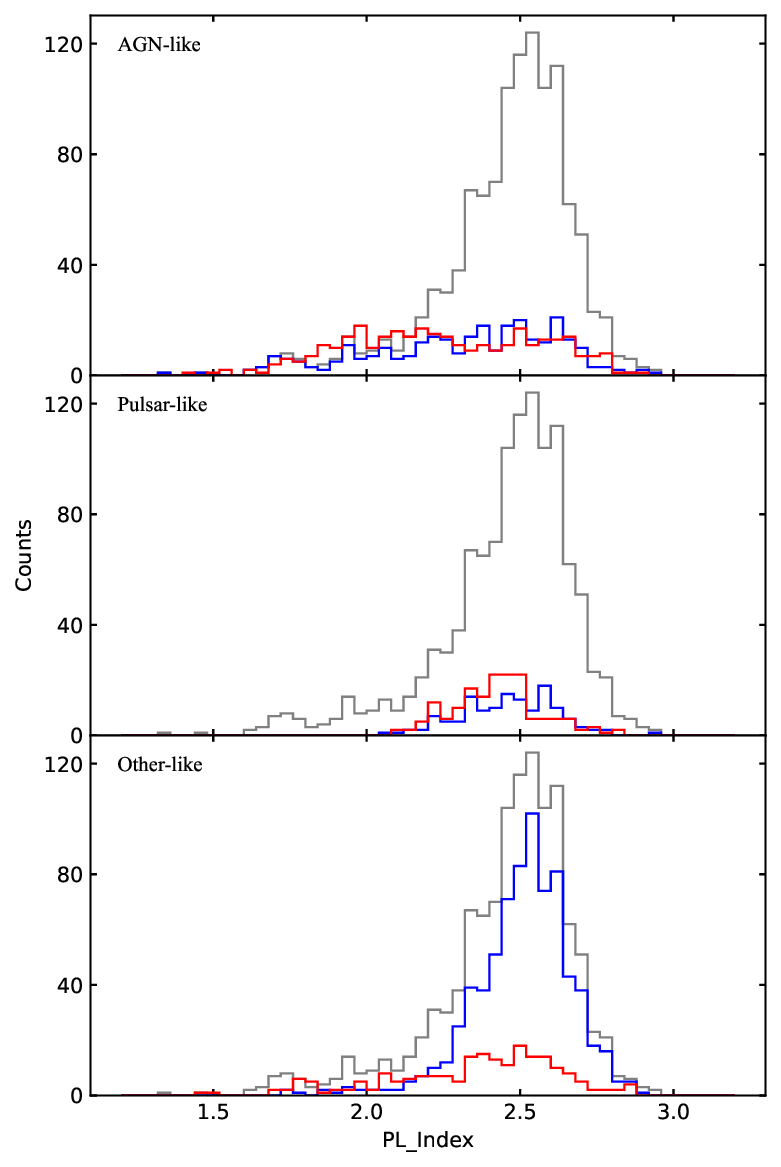}
\caption{The gamma-ray spectral index distribution of low-latitude unassociated sources, associated sources, and the candidates identified by the ML. The gray line represents all the unassociated sources, the red color represents associated sources of different categories, and the blue color represents candidates of different categories identified by ML.}
 \label{fig6}
\end{figure}

We investigated the spectral index distributions of the candidate sources, with different candidate categories shown in blue in Figure \ref{fig6}. From the figure, it can be seen that the LGL unassociated sources are predominantly other-like, and their spectral index distribution is similar to that of the unassociated sources (gray line). However, the AGN-like candidates still exhibit an excessive soft component, which is not reasonable. Additionally, we obtained 290 AGN candidate sources, and when combined with the existing 326 high-confidence associated AGN samples, the total number of AGNs reached 616, significantly exceeding the estimated value of $452 \pm 27$. The higher density of AGNs in the low latitude region compared to the high latitude region is clearly unreasonable.

We re-evaluated the AGN-like candidates using a BG model, as shown in Figure 3. Among them, 83 candidates were identified as non-AGN-like and were labeled as low-confidence AGN-like candidates (LACs), while the remaining 207 samples were considered high-confidence AGN-like candidates (HAC). By combining the HACs with the 326 high-confidence associated AGN samples,  shown as the blue area at the bottom panel of Figure \ref{fig3}, the excess of soft spectral sources was suppressed, resulting in a total of 533 sources, slightly higher than the estimated number, which can be considered a reasonable outcome.

\section{Results } \label{sec:Result}

\begin{figure*}
\centering
\includegraphics[height=7.4cm,width=12cm]{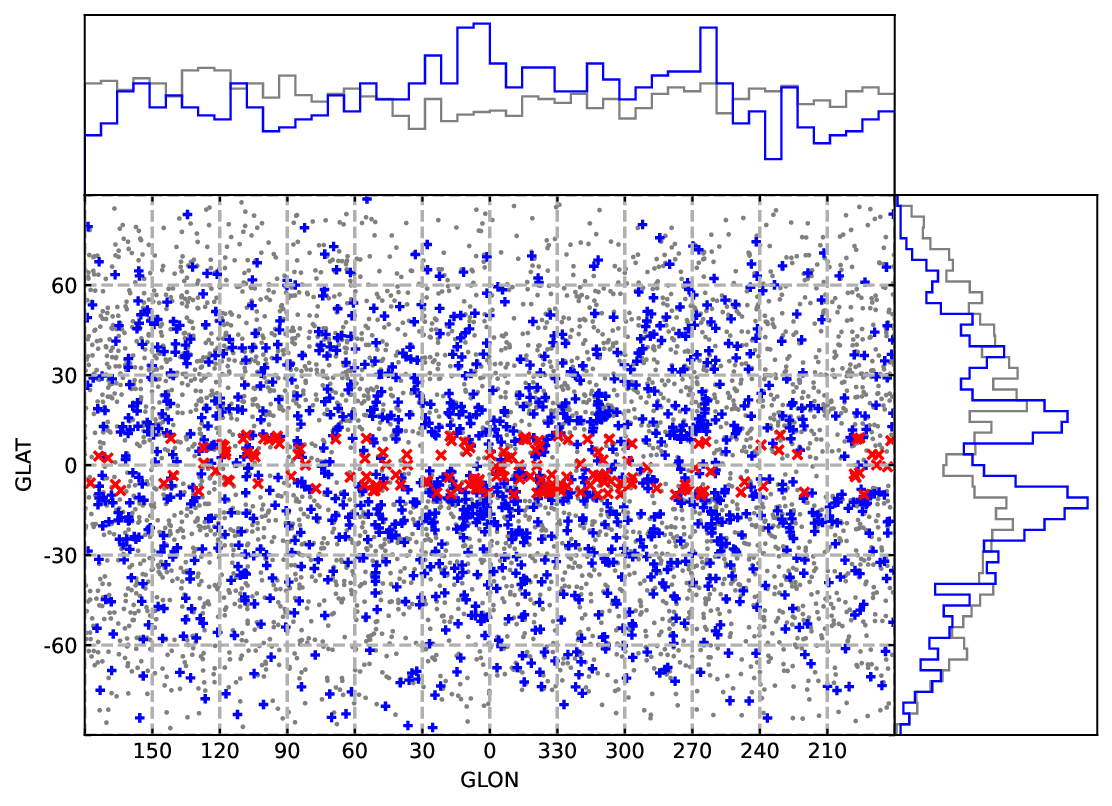}
 \includegraphics[height=7.4cm,width=12cm]{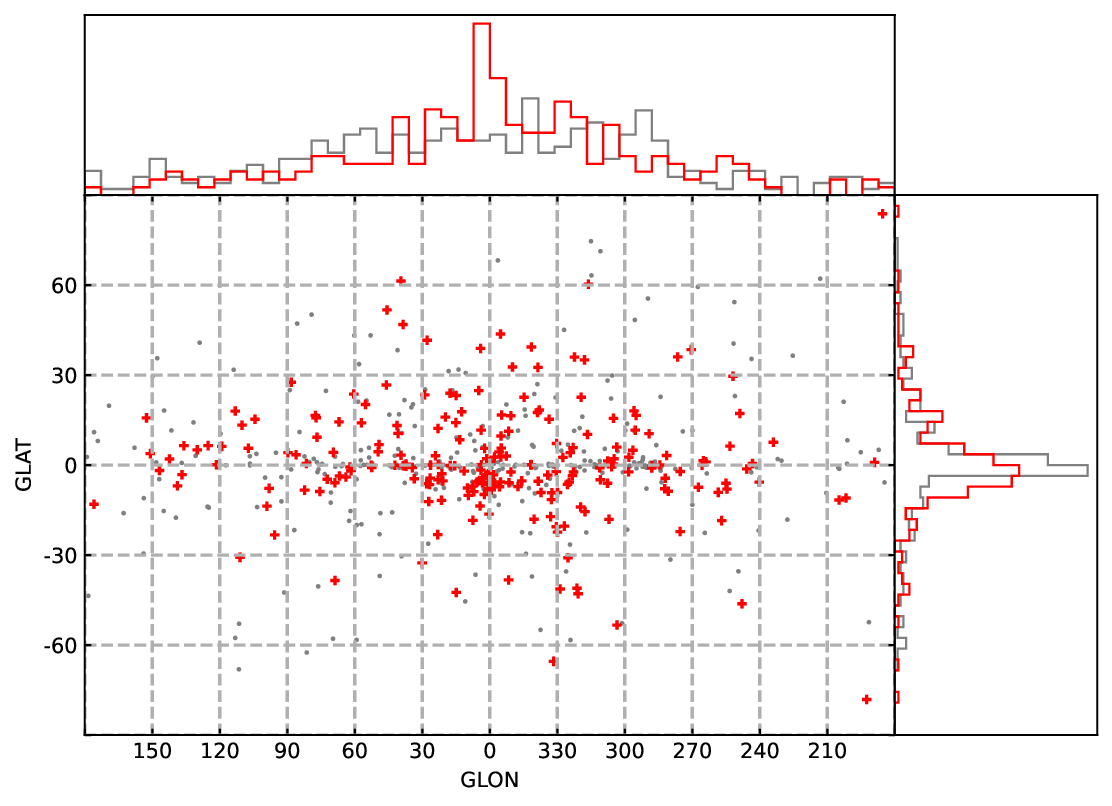}
  \includegraphics[height=7.4cm,width=12cm]{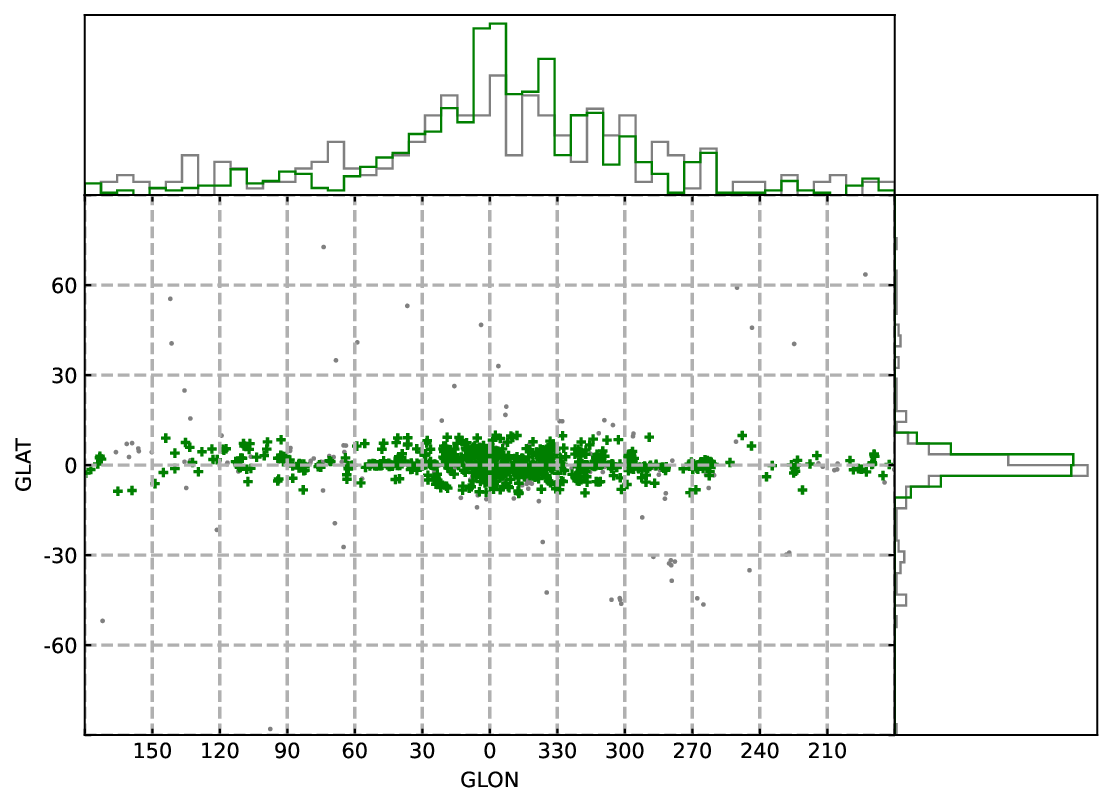}
\caption{All-sky scatter plot and density distribution of associated and candidates of AGN-like, Pulsar-like, and other-like sources. The upper panel shows the AGN category, with candidates represented in green, associated sources in gray, and low-confidence samples removed by the BG model in red. The middle panel displays the pulsar category, with candidates shown in red and associated sources in gray. The lower panel presents the other-like category, with candidates depicted in green and associated sources in gray. The density distribution curves provide insights into the spatial distribution of unassociated sources across Galactic latitude and longitude.}
\label{fig7}
\end{figure*}

In the previous sections, we conducted models for high-latitude and low-latitude sources and classified the unassociated sources in 4FGL-DR3. Combining the classification results from these two frameworks, we constructed an all-sky catalog of unassociated sources. Among the 2291 unassociated sources, 1327 were identified as AGN-like candidates (1244 HAC, 83 LAC), 223 as Pulsar-like candidates, and 741 as Other-like candidates. In Figure \ref{fig7}, we presented the scatter plot of the associated sources and candidates in the Galactic coordinate system, as well as the density distribution curves for Galactic longitude and Galactic latitude.

In the upper panel of Figure \ref{fig7}, we presented the plot related to AGN. The gray dots represent known AGN samples, the blue dots represent AGN candidates identified through LGL classification, and the red dots represent LACs. Associated active galactic nuclei are widely distributed throughout the celestial sphere, but there is a gap near the Galactic plane. A large number of sources cluster around the Galactic center, especially among the unassociated sources, resulting in a concentration of candidates identified by ML in that region. Furthermore, considering the strong diffuse gamma-ray background near the Galactic center, such a concentration have intensified. Applying the Bayesian-Gaussian model for correction, the excluded samples are mainly concentrated near the Galactic center. However, even after the correction, the remaining AGN-like candidates (HACs) still exhibit a distribution near the Galactic center at Galactic longitude that exceeds the expected. This indicates that our constraints are still insufficient, and there is still a part of non-AGN contamination among the AGN-likes. Additionally, this was consistent with the total number of associated AGNs and AGN candidates exceeding the expected limit.

The middle panel of the figure corresponds to the Pulsar plot. The gray dots represent known pulsar samples, and the red dots represent pulsar-like candidates identified through ML. The number of high Galactic latitude pulsar-like candidates is slightly lower than that of LGL candidates. Comparing the distribution of associated pulsars and pulsar-like candidates, we can observe an accumulation of pulsar-like candidates near the Galactic center. However, due to the strong gamma-ray background near the Galactic center, the validity of these samples needs to be carefully considered.

The lower panel of figure \ref{fig7} represents the Other-like plot. The gray dots represent known samples, and the green dots represent other-like candidates identified through machine learning. Since the high Galactic latitude region was classified using an AGN-pulsar binary classifier, the resulting other-like candidates are only present in the low Galactic latitude region. The Other category is more complex as it includes both Galactic components (PWN, SNR, etc.) and extragalactic components (galaxies, etc.). Both the candidates and associated sources exhibit a symmetric distribution centered around the celestial coordinates. It should be noted that the AGN-like candidates excluded using the Bayesian-Gaussian model are considered as other-like samples, but they are not depicted in the plot.

We present a machine learning classification catalog of unassociated sources, which consists of the following three parts:

\begin{itemize}
\item [1.]
High Galactic Latitude Unassociated Source Catalog (una$\_$high.fits): This catalog includes source information for HGL unassociated sources along with their ML classification results.
\item [2.]
Low Galactic Latitude Unassociated Source Catalog (una$\_$low.fits): This catalog contains source information for LGL unassociated sources, their ML classification results, and the re-evaluated results using the BG model.
\item [3.]
Misassociated Low Galactic Latitude AGN Candidate Catalog (agn$\_$low.fits): This catalog provides source information for 80 LGL samples that bg model considers unlikely to be AGNs. It also includes the likelihood probabilities of being AGN/non-AGN based on the Bayesian Gaussian model.
\end{itemize}

All tables are available in FITS format and  can be accessed online through the supplementary material provided by MNRAS (See Data Availability).  Detailed descriptions of each column in these three FITS tables can be found in Table \ref{Tab4}.

\section{Conclusion and Discussion } \label{sec:Conclu}

In this paper, we divided the task of classifying unassociated of Fermi-LAT gamma-rays sources into two frameworks.

In the high Galactic latitude region, we employed a binary classifier to classify AGN-like and Pulsar-like sources and trained it using imbalanced samples. By utilizing four supervised machine learning algorithms and optimizing the models, we achieved a balanced accuracy of 90$\%$ in a 5-fold stratified cross-validation experiment. The predicted results exhibited consistency. The classification results obtained from the four algorithms demonstrated a high level of consistency. By employing an ensemble voting classifier, we identified 1037 AGN candidates and 88 pulsar candidates with a balanced accuracy of $0.918 \pm 0.029$.

In the low Galactic latitude region, the number of unassociated sources exceeded the number of associated sources, and the features of the sources were not clear, resulting in challenges in the classification process.
We introduced BG model by fitting Gaussian functions to the distributions of gamma-ray spectral index, variability index, and log-parabolic fit significances of the associated sources.
During the evaluation of associated AGNs, we identified 81 sources with low confidence as misassociated candidates, which is consistent with the findings of \cite{2022ApJS..260...53A}.
After removing these samples from the training set, we constructed a three-class classifier for AGN-like, pulsar-like, and other-like sources using the same four supervised ML algorithms.
The balanced accuracies of the various classifiers in the three-class classification were all close to 80$\%$. By employing an ensemble voting classifier, we obtained 290 AGN-like candidates, 135 pulsar-like candidates, and 742 other-like candidates, achieving a balanced accuracy of $0.815 \pm 0.027$.
 After re-evaluating the AGN-like candidates using the BG parameter model, we found 83 candidates with low confidence, label as LAC. Our ML results directly indicated that non-AGN and non-pulsar sources dominate the low Galactic latitude region.

By combining the classification results obtained from the high and low Galactic latitude regions, we constructed a comprehensive catalog of unassociated sources across the all-sky.

In the HGL region, a simple method achieved a high accuracy on the training set, and no significant anomalies were found when examining the parameter space of the classification results. However, several challenges were encountered in the LGL region. Firstly, the number of unassociated sources exceeds the number of associated sources,  posing challenges to model construction; Secondly, the purity of the training samples is not high, with an excessive number of soft spectrum samples in the LGL region, raising suspicion of contamination from other sources. Thirdly, there is a significant mismatch in the proportions of different categories between the training samples and the samples for predicted. In the training set, the ratios of AGN-like, Pulsar-like, and other-like sources are 407:166:209, while based on our classification results, the ratios of these three categories are 290:135:742, indicating a severe deviation between these two ratios. All of these factors pose challenges to the performance of machine learning classifiers, and the generalization ability from the training set to the test set is questionable. Additionally, the accuracy of the classifiers in the low Galactic latitude region is lower compared to the high Galactic latitude region, and there are differences in the results among different classification algorithms, especially for the pulsar category, resulting in fluctuations in the predictions for unassociated sources, ranging from 109 to 199.

We compared the classification results of our work with the results from early attempts, such as  \cite{2021RAA....21...15Z}, \cite{2022A&A...660A..87B}, and \cite{2022MNRAS.515.1807C} , and the specific details are listed in Appendix \ref{append:1}.  The results show that current ML methods have achieved high accuracy on the all-sky training set. Successful classification of unassociated sources in the HGL region has reached a consensus: these sources are predominantly dominated by AGNs, with a small fraction being pulsar-likes and other-likes categories. However, the all-sky ML models face challenges in classifying unassociated sources in the LGL region, and there is considerable inconsistency among the results of different classifiers. In single-classifier classification, the predicted results indicate that unassociated sources near the Galactic plane are dominated by AGNs. In multi-classifier classification, although the results are unified through ``All-Agree", more than half of the low-Galactic-latitude sources are classified as ``MIXED" without specific classes.

Although our results directly indicated that unassociated sources in the low-latitude region are neither dominated by AGNs nor pulsars, this is consistent with the distribution of these sources in parameter space, such as gamma-ray spectral index. However, the ``other-like" category is more complex, including both Galactic and extragalactic components, and the sample size for each individual category is small (the maximum is 114). This poses difficulties in constructing machine learning models for further classification, which are more suitable for large-scale data mining and analysis. Additionally, according to \cite{2022ApJS..260...53A}, unassociated sources near the Galactic plane (Gus, $-3^{\circ}<b<3^{\circ}$) have been occupying an increasing proportion in recent years and exhibit an unusually dense distribution. Furthermore, a special distribution is observed in a subset of LGL unassociated sources characterized by extreme softness (SGUs, $\Gamma>2.4$), with no known Galactic gamma-ray emitting sources similar to them. Currently, it is speculated that residual background or undiscovered new gamma-ray sources contribute to this phenomenon. Our study has largely ruled out the association of these sources with AGN and pulsars, but there is still a significant gap in achieving detailed classification.

In this study, we used the BG model to filter out soft-spectrum excessed samples from LGL AGN cores and removed them from the training and testing sets of the LGL classifier. To validate the effectiveness of removing abnormal samples using the BG parameter model, we compared the models trained on different datasets,  including datasets with and without the removal of soft-spectrum abnormal AGN and their incorporation into the Other-like category.  The specific results can be found in Appendix \ref{append:2}. The results showed that after removing 81 soft-spectrum abnormal AGN from the training set, the test balanced accuracy of various classification models significantly improved, yielding stable results. However, the obtained AGN-like candidate samples still included some soft-spectrum abnormal samples. Furthermore, the results indicated that minor changes in the training set would not significantly affect the accuracy of the classifier but would completely alter the classification predictions.

Previous research has explored various methods for handling imbalanced sample classification, such as adjusting training sample weights and oversampling techniques like Synthetic Minority Over-sampling Technique (SMOTE), in the context of classification e.g., \citealt{2021RAA....21...15Z,2022A&A...660A..87B}). However, these studies primarily focuses on evaluating the performance of classifiers, such as accuracy, on the training and test sets, without discussing the impact on prediction results. In this study, we investigate the effects of adjusting model hyper-parameters and applying the SMOTE algorithm for oversampling on the accuracy of a binary classifier for unassociated sources at high Galactic latitudes. We also provide evaluation results of different models for high Galactic latitude unassociated sources. Detailed information can be found in Appendix \ref{append:3}. The results show that artificially changing the weights of training samples does not decrease the accuracy of the classifier, and may even improve it. However, significantly alters the prediction results.

Previous investigations have delved into a multitude of methodologies to address imbalanced sample classification. However, these studies primarily focused on assessing classifier performance using metrics such as accuracy on training and test sets, without considering the impact on prediction outcomes. We have thoroughly examined the effects of adjusting model hyper-parameters and employing the SMOTE algorithm for oversampling on the accuracy of a binary classifier dedicated to high Galactic latitude unassociated sources in this work. Additionally, we have presented evaluation results for diverse models targeting high Galactic latitude unassociated sources. For comprehensive details, please refer to Appendix \ref{append:3}. The outcomes reveal that modifying the weights of training samples artificially does not diminish classifier accuracy; in fact, it may even enhance it (refer to Table \ref{Tabc1}). Nevertheless, this practice significantly alters the prediction results and leads to more substantial disparities in classification outcomes among different classifiers. Thus, further scrutiny is imperative to ascertain the reliability of this approach.

Because of the unique properties of astronomical data, sources with higher significance are usually detected first, while sources with lower significance are more difficult to identify. In high Galactic latitude regions, the detection of samples is relatively easier, whereas in the Galactic plane, the high source density and strong background radiation pose challenges for detection. This evident imbalance challenges the assumption of sample representativeness in machine learning. This problem is not only present in the classification of unassociated sources in Fermi-LAT but also in various other astronomical applications of ML.

Based on the above analysis, we provide the following recommendations for ML classification of  Fermi-LAT unassociated sources:
\begin{itemize}
 \item [1.]
The high Galactic latitude classification has been successful, so it is advisable to focus more on the low Galactic latitude region for further improvement.
 \item [2.]
Considering the systematic differences between unassociated sources and associated sources, it is important to prioritize the rationality of classification results when constructing the model, rather than solely focusing on the classifier's performance on known samples. In such an imperfect dataset, the accuracy of the classifier alone cannot fully represent its performance. It is necessary to involve a reasonable assessment of the predictive capability of the samples.
\end{itemize}

This study divided the Galactic latitude region into high and low regions using a threshold of $|b|=10^\circ$. However, it is worth noting that there is a significant enrichment of sources in the region where $10^\circ<b<20^\circ$ within the range of mid-to-low Galactic latitude range, and previous classification results have also shown considerable differences. Additionally, although we partially considered the differences between high and low Galactic latitude samples, we did not account for variations in significance levels and variability between high and low classes. This aspect will be considered in future work.

\begin{landscape}
\begin{table}
\centering
\caption{Detailed information of each column in the results table}\label{Tab4}
\resizebox{23cm}{!}{
\begin{tabular}{ccccccc}
\hline \hline
\multirow{2}{*}{Label}&\multirow{2}{*}{Column name}& \multirow{2}{*}{Description}&&\multirow{2}{*}{Label}&\multirow{2}{*}{Column name}& \multirow{2}{*}{Description}\\
&&&\\
\normalsize(1) & \normalsize(2) & \normalsize(3) &&\normalsize(1) & \normalsize(2) & \normalsize(3) \\
\hline
\multicolumn{3}{c}{una$\_$low.fits}&&\multicolumn{3}{c}{una$\_$high.fits}\\	
\cline{1-3}\cline{5-7}
1&Source$\_$Name&Source Name of LGL unassociated sources from 4FGL-DR3&&1&Source$\_$Name&Source Name of HGL unassociated sources from 4FGL-DR3\\
2&Class$\_$FGL&High-confidence source categories from 4FGL-DR3&&2&Class$\_$FGL&High-confidence source categories from 4FGL-DR3\\
3&GLON &\multirow{3}{*}{Source feature parameters in Table \ref{Tab1}}&&3&GLON&\multirow{3}{*}{Source feature parameters in Table \ref{Tab1}}\\
-& - &&&-& - &\\
42&H678&&&42&H678&\\
\multirow{2}{*}{43}&\multirow{2}{*}{pro$\_$agn$\_$lr}&The likelihood probability assigned by single LR classifier &&\multirow{2}{*}{43}&\multirow{2}{*}{pro$\_$agn$\_$lr}&The likelihood probability assigned by single LR classifier \\
&&that the source belongs to the "AGN-like"&&&&that the source belongs to the "AGN-like"\\
\multirow{2}{*}{44}&\multirow{2}{*}{pro$\_$psr$\_$lr}&The likelihood probability assigned by single LR classifier &&\multirow{2}{*}{44}&\multirow{2}{*}{pro$\_$psr$\_$lr}&The likelihood probability assigned by single LR classifier \\
&&that the source belongs to the "Pulsar-like"&&&&that the source belongs to the "Pulsar-like" \\
45&class$\_$lr&The evaluated unassociated source categories from single LR classifier &&45&class$\_$lr&The evaluated unassociated source categories from single LR classifier \\
\multirow{2}{*}{46}&\multirow{2}{*}{pro$\_$agn$\_$svm}&The likelihood probability assigned by single SVM classifier &&\multirow{2}{*}{46}&\multirow{2}{*}{pro$\_$agn$\_$svm}&The likelihood probability assigned by single SVM classifier \\
&&that the source belongs to the "AGN-like"&&&&that the source belongs to the "AGN-like"\\
\multirow{2}{*}{47}&\multirow{2}{*}{pro$\_$psr$\_$svm}&The likelihood probability assigned by single SVM classifier &&\multirow{2}{*}{47}&\multirow{2}{*}{pro$\_$psr$\_$svm}&The likelihood probability assigned by single SVM classifier \\
&&that the source belongs to the "Pulsar-like"&&&&that the source belongs to the "Pulsar-like" \\
48&class$\_$svm&The evaluated unassociated source categories from single SVM classifier &&48&class$\_$svm&The evaluated unassociated source categories from single SVM classifier \\
\multirow{2}{*}{49}&\multirow{2}{*}{pro$\_$agn$\_$rf}&The likelihood probability assigned by single RF classifier &&\multirow{2}{*}{49}&\multirow{2}{*}{pro$\_$agn$\_$rf}&The likelihood probability assigned by single RF classifier \\
&&that the source belongs to the "AGN-like"&&&&that the source belongs to the "AGN-like"\\
\multirow{2}{*}{50}&\multirow{2}{*}{pro$\_$psr$\_$rf}&The likelihood probability assigned by single RF classifier &&\multirow{2}{*}{50}&\multirow{2}{*}{pro$\_$psr$\_$rf}&The likelihood probability assigned by single RF classifier \\
&&that the source belongs to the "Pulsar-like"&&&&that the source belongs to the "Pulsar-like" \\
51&class$\_$rf&The evaluated unassociated source categories from single RF classifier &&51&class$\_$rf&The evaluated unassociated source categories from single RF classifier \\
\multirow{2}{*}{52}&\multirow{2}{*}{pro$\_$agn$\_$mlp}&The likelihood probability assigned by single MLP classifier &&\multirow{2}{*}{52}&\multirow{2}{*}{pro$\_$agn$\_$mlp}&The likelihood probability assigned by single MLP classifier \\
&&that the source belongs to the "AGN-like"&&&&that the source belongs to the "AGN-like"\\
\multirow{2}{*}{53}&\multirow{2}{*}{pro$\_$psr$\_$mlp}&The likelihood probability assigned by single MLP classifier &&\multirow{2}{*}{53}&\multirow{2}{*}{pro$\_$psr$\_$mlp}&The likelihood probability assigned by single MLP classifier \\
&&that the source belongs to the "Pulsar-like"&&&&that the source belongs to the "Pulsar-like" \\
54&class$\_$mlp&The evaluated unassociated source categories from single MLP classifier &&54&class$\_$mlp&The evaluated unassociated source categories from single MLP classifier \\
\multirow{2}{*}{55}&\multirow{2}{*}{pro$\_$agn$\_$vote}&The likelihood probability assigned by ensemble voting classifier &&\multirow{2}{*}{55}&\multirow{2}{*}{pro$\_$agn$\_$vote}&The likelihood probability assigned by ensemble voting classifier \\
&&that the source belongs to the "AGN-like"&&&&that the source belongs to the "AGN-like"\\
\multirow{2}{*}{56}&\multirow{2}{*}{pro$\_$psr$\_$vote}&The likelihood probability assigned by ensemble voting classifier &&\multirow{2}{*}{56}&\multirow{2}{*}{pro$\_$psr$\_$vote}&The likelihood probability assigned by ensemble voting classifier \\
&&that the source belongs to the "Pulsar-like"&&&&that the source belongs to the "Pulsar-like" \\
57&class$\_$vote&The evaluated unassociated source categories from ensemble voting classifier &&57&class$\_$vote&The evaluated unassociated source categories from ensemble voting0 classifier \\
\multirow{2}{*}{58}&\multirow{2}{*}{pro$\_$agn$\_$bg}&The likelihood probability assigned by BG model&& \\
&&that the ML-selected AGN-like candidates belongs to the AGN&&&&\\
\multirow{2}{*}{59}&\multirow{2}{*}{pro$\_$non$\_$agn$\_$bg}&The likelihood probability assigned by BG model && \\
&&that the ML-selected AGN-like candidates belongs to the non-AGN&&&& \\
60&class$\_$bg&The evaluated class of ML-selected AGN-like candidates from BG model &&\\
60&flag$\_$bg&The confidence flag of AGN-like candidates combining ML classifiers and the BG model&&\\

\hline
\multicolumn{3}{c}{agn$\_$low.fits}\\	
\cline{1-3}
1&Source$\_$Name&Source Name of LGL AGNs from 4FGL-DR3&&\\
2&GLON&Galactic longitude&&\\
3&GLAT&Galactic latitude&&\\
4&Class$\_$FGL&Source class from 4FGL-DR3&&\\
5&PL$\_$Index&The parameter PL$\_$Index used in the BG model&&\\
6&log(LP$\_$SigCurv)&The parameter log(LP$\_$SigCurv) used in the BG model&&\\
7&log(Variability$\_$Index)&The parameter log(Variability$\_$Index) used in the BG model&&\\
\multirow{2}{*}{8}&\multirow{2}{*}{pro$\_$agn$\_$bg}&The likelihood probability assigned by BG model&&\\
&&that the LGL AGNs belongs to the AGN&&\\
\multirow{2}{*}{9}&\multirow{2}{*}{pro$\_$non$\_$agn$\_$bg}&The likelihood probability assigned by BG model&&\\
&&that the LGL AGNs belongs to the non-AGN&&\\
10&class$\_$bg&The evaluated class of LGL AGNs from BG model&&\\
\hline
\end{tabular}}\\
\end{table}
\end{landscape}

\section*{Acknowledgements}
We thank the anonymous referee for very constructive and helpful comments and suggestions, which greatly helped us to improve our paper.
We would like to thank the Fermi Science Support Center (FSSC) for the public availability of data.
This work is partially supported by the National Natural Science Foundation of China NSFC 12233006 and the Yunnan University graduate research innovation fund project.

{\it Facility:} Fermi-LAT

{\it Software:}  scikit-learn \citep{scikit-learn}, Matplotlib \citep{Hunter:2007}, NumPy \citep{ harris2020array}, Astropy \citep{astropy:2022}, SciPy \citep{2020SciPy-NMeth}, imbalanced-learn \citep{lemaitre2017imbalanced}.

\section*{Data availability}
The results of this study are available in FITS format as supplementary material online from MNRAS.

\bibliographystyle{mnras}
\bibliography{article}

\appendix

\section{Comparison of machine learning classification results in recent years}  \label{append:1}

\begin{figure*}
\centering
\begin{adjustbox}{valign=T}
\includegraphics[height=11.5cm,width=8cm]{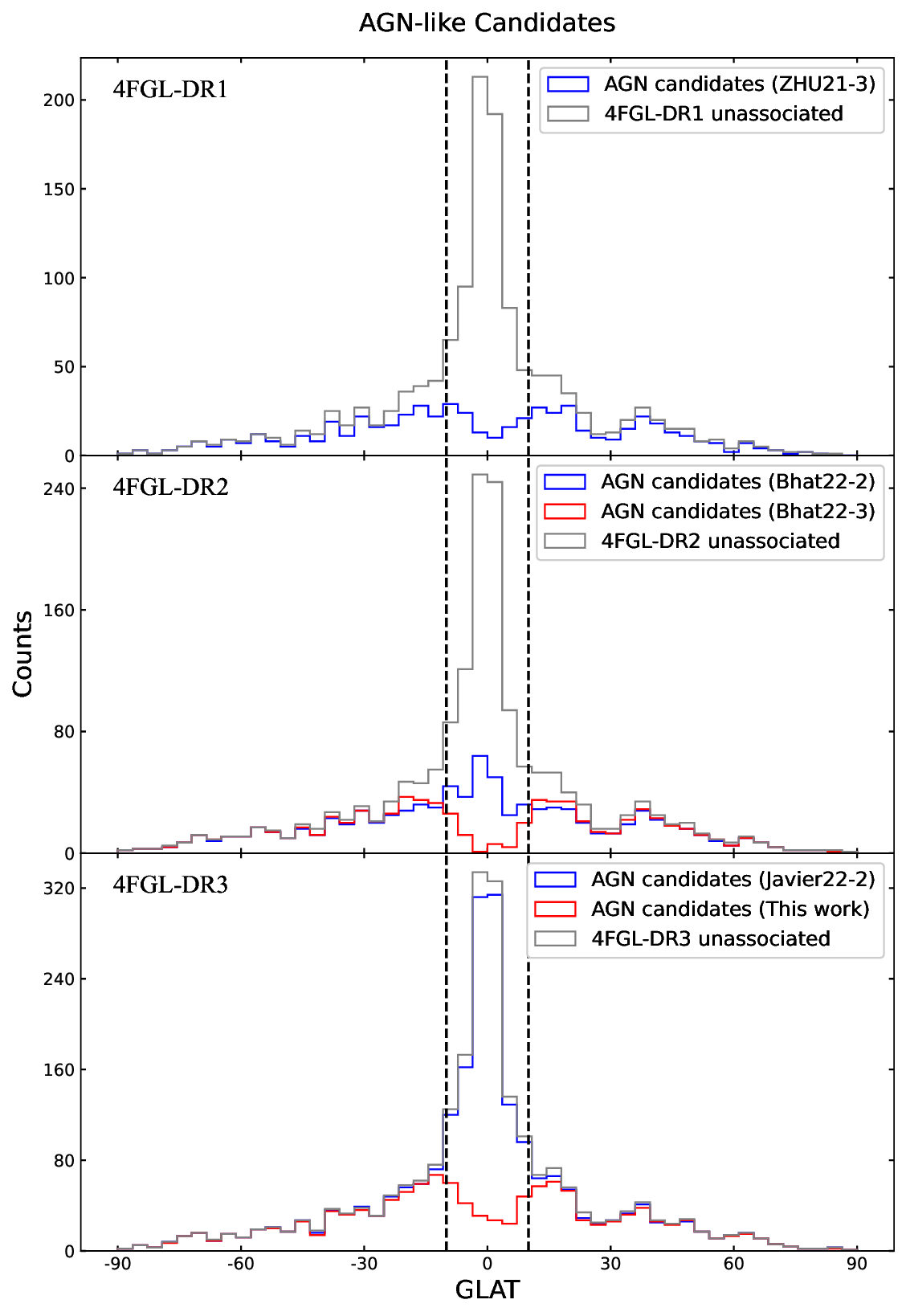}
\end{adjustbox}
\begin{adjustbox}{valign=T}
\includegraphics[height=11.5cm,width=8cm]{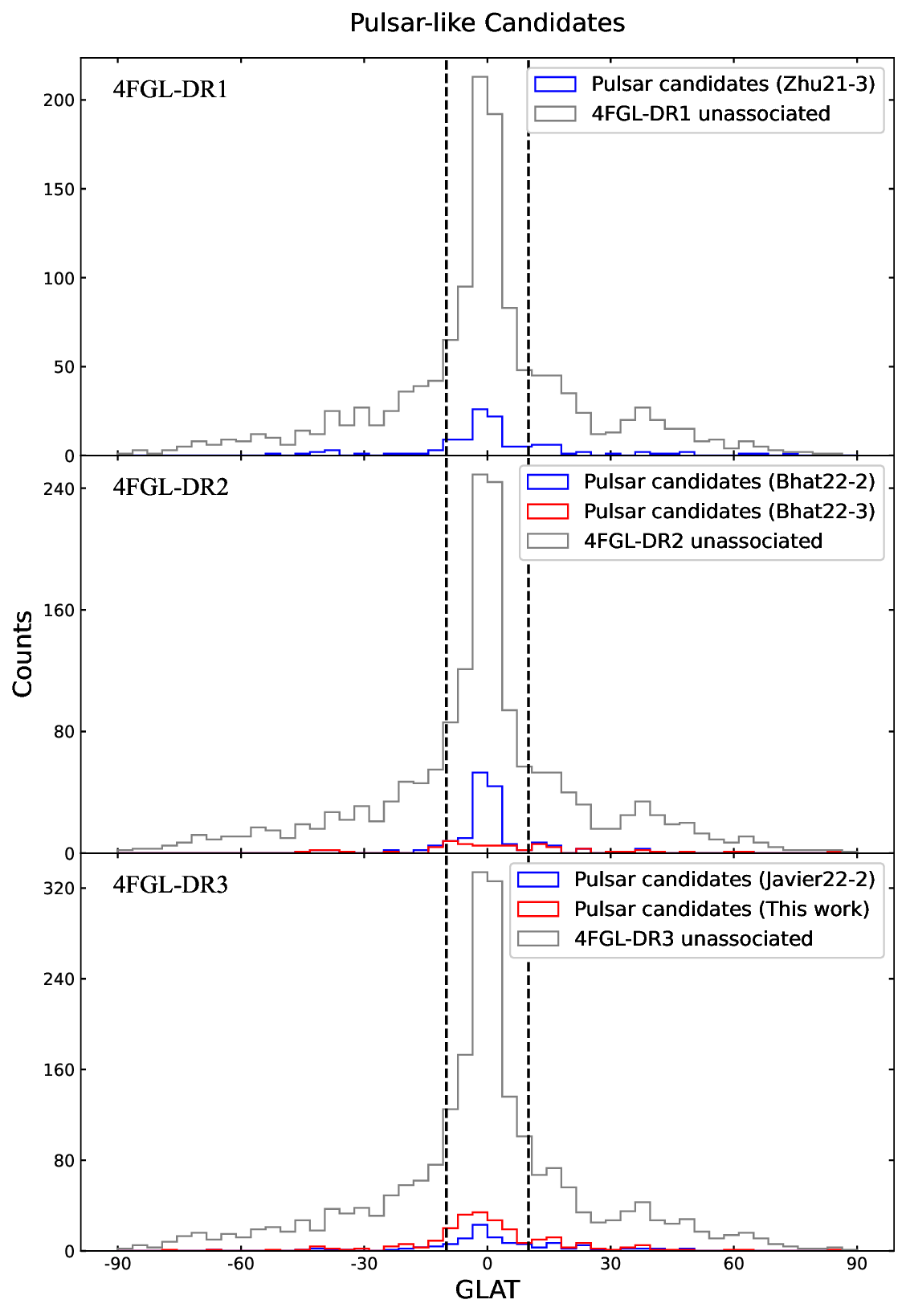}
\end{adjustbox}
\begin{adjustbox}{valign=T}
\includegraphics[height=11.5cm,width=8cm]{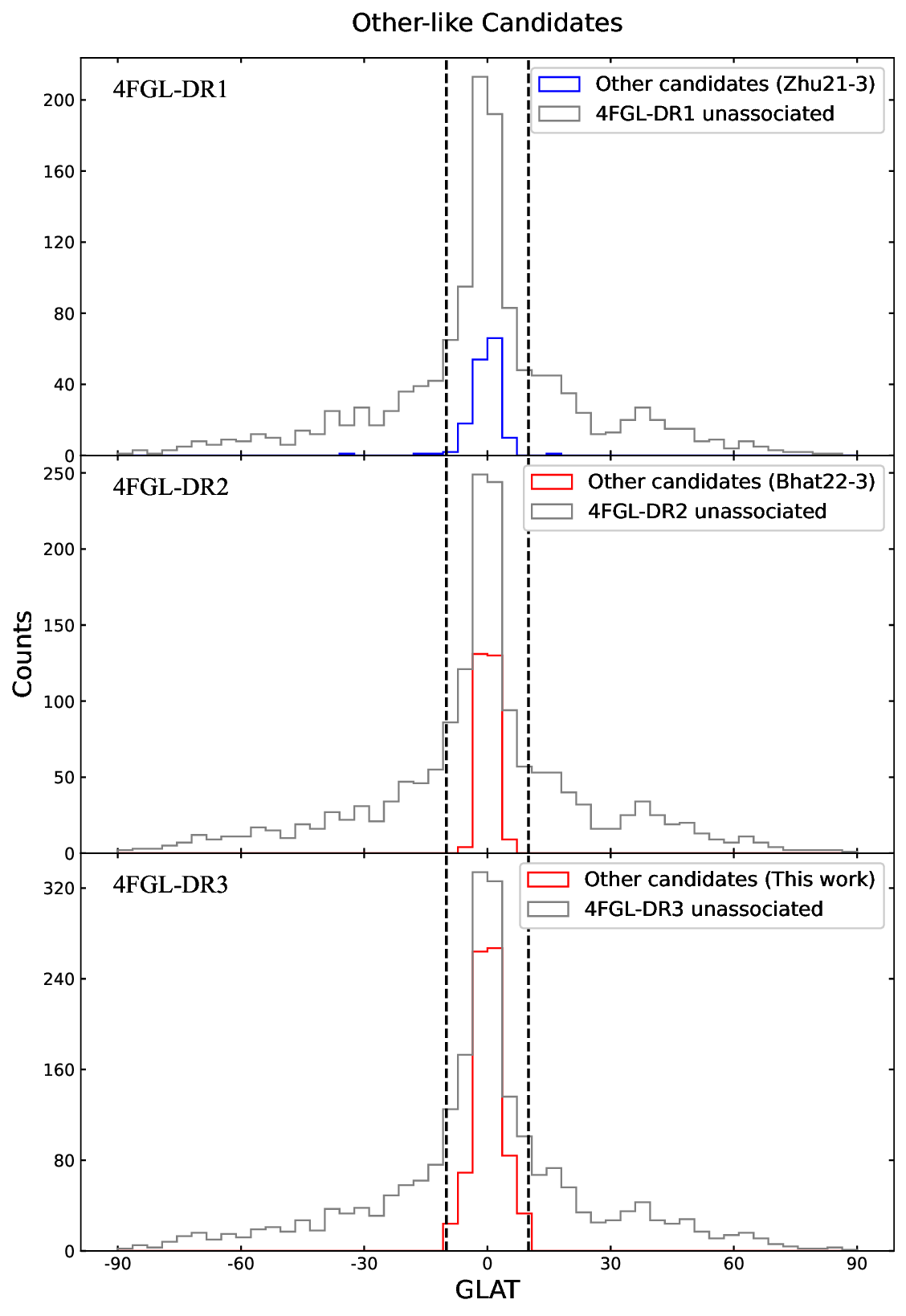}
\end{adjustbox}
\begin{adjustbox}{valign=T}
\includegraphics[height=8cm,width=8cm]{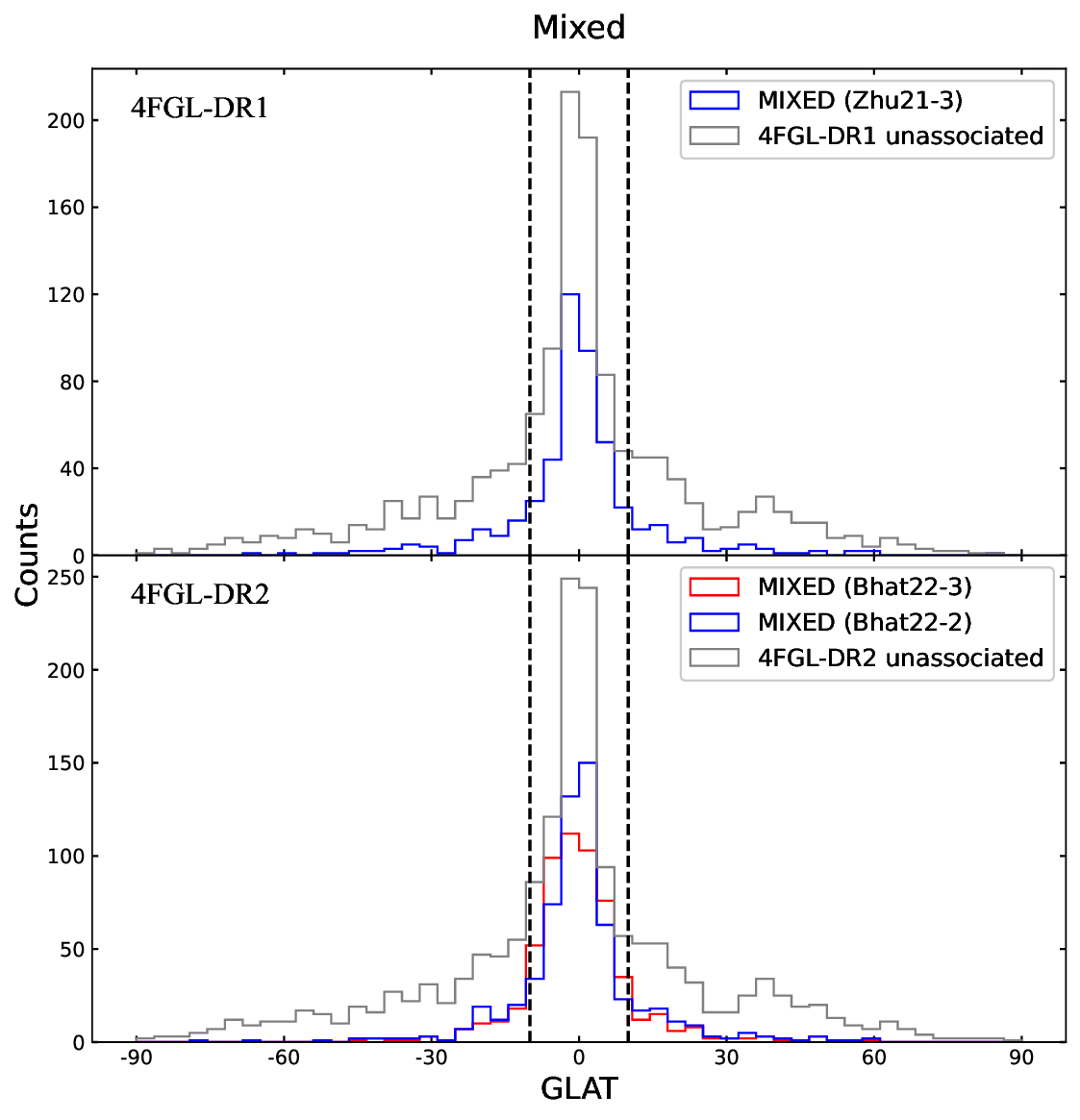}
 \end{adjustbox}
\caption{The density Distribution of Classification Results in Galactic Latitude}
\label{figA1}
\end{figure*}

ML has been widely applied in the classification of unassociated sources in Fermi-LAT. We collected classification results from \cite{2021RAA....21...15Z}, \cite{2022A&A...660A..87B}, and \cite{2022MNRAS.515.1807C}. In \cite{2021RAA....21...15Z}, RF and Neural Network (NN) were used for classification. Initially, a ternary classification of AGN-like, PSR-like, and Other-like sources was performed, followed by combining the classification results using the ``All-agree" method to classify unassociated sources in 4FGL-DR1 (labeled as Zhu21-03). In \cite{2022A&A...660A..87B}, RF, Boosted Decision Trees (BDT), NN, and LR were used, and the ``SMOTE" algorithm was applied for oversampling. Eight classifiers were used for binary classification of AGN-like and PSR-like sources (labeled as Bhat22-02), as well as ternary classification of AGN-like, PSR-like, and Other-like sources (labeled as Bhat22-03). The classification results were combined using the ``All-agree" method to construct a probabilistic source catalog. In \cite{2022MNRAS.515.1807C}, a CATBOOST algorithm was used to build a all-sky model for binary classification (labeled as Javier22-02) and multi-class classification of unassociated sources in 4FGL-DR3.

\begin{figure*}
\centering
\begin{adjustbox}{valign=T}
\includegraphics[height=8cm,width=8cm]{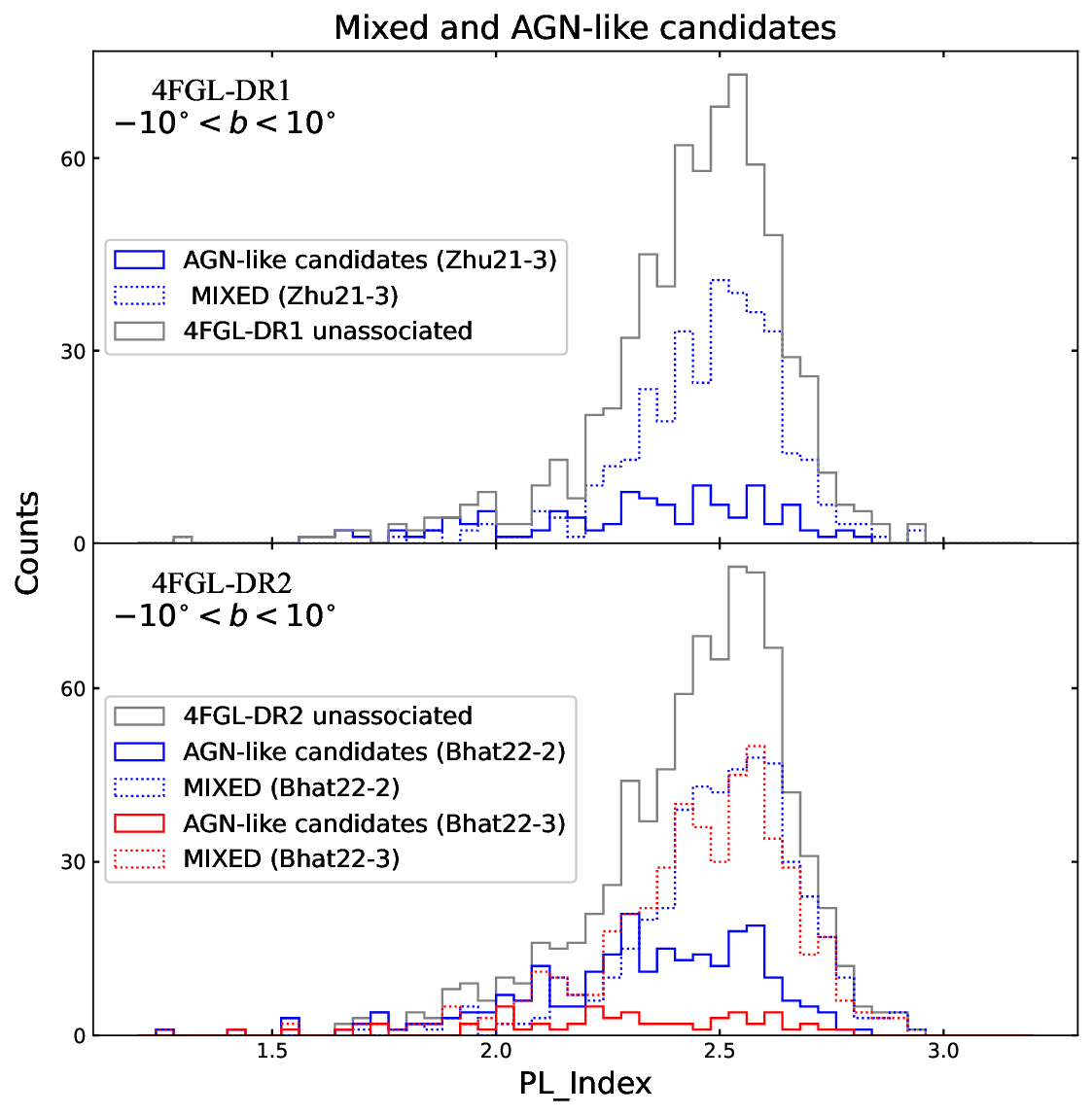}
\end{adjustbox}
\begin{adjustbox}{valign=T}
\includegraphics[height=11.5cm,width=8cm]{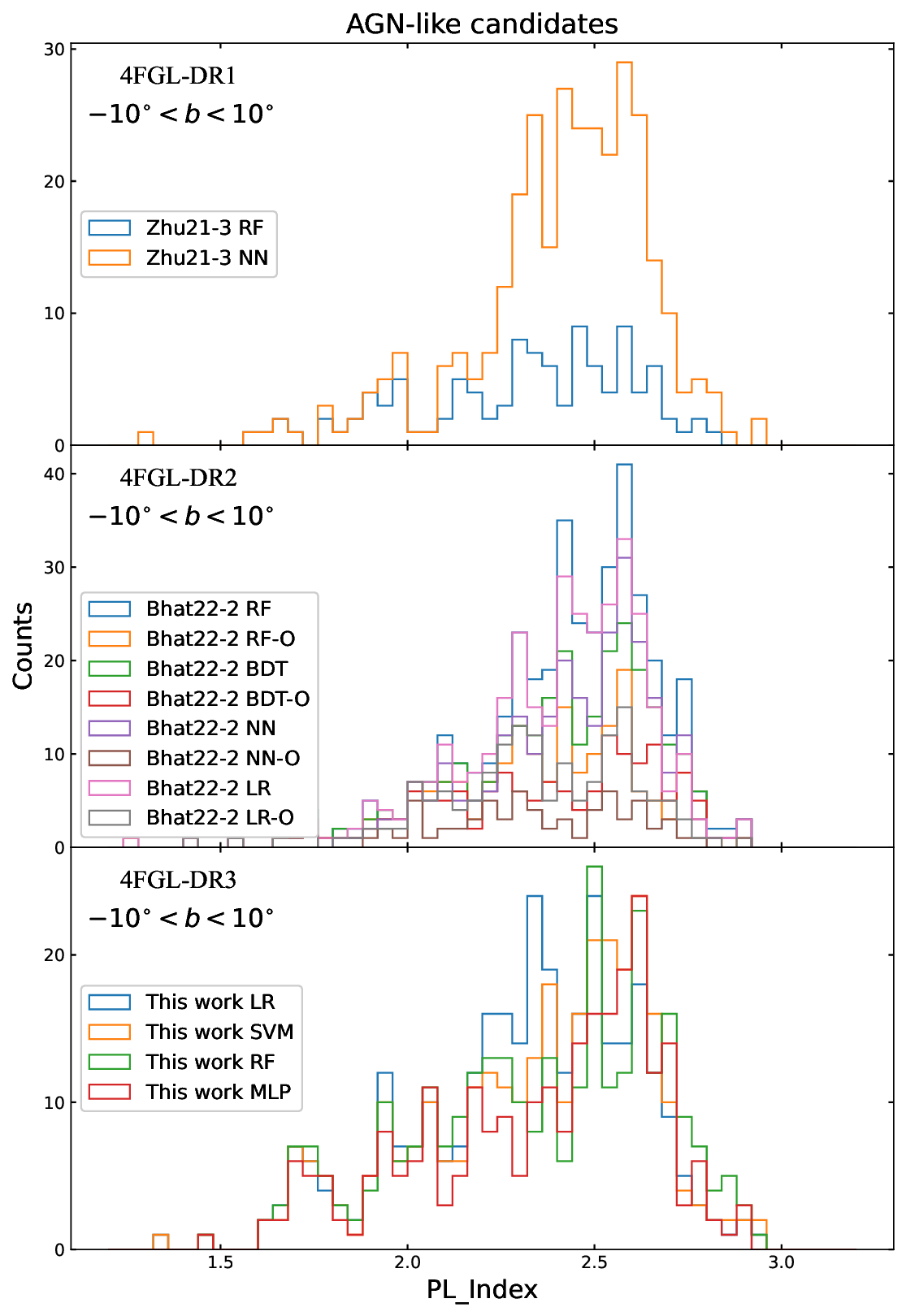}
\end{adjustbox}
\caption{The density Distribution of ``MIXED" class and AGN-like candidates in Gamma-ray Spectral Index}
\label{figA2}
\end{figure*}

Combining these results with our own, Figure \ref{figA1} shows the density distribution of the classification results in Galactic latitude. It is evident from Figure \ref{figA1} that the current attempts have achieved consistent results in the HGL region for both binary classification of AGN-Pulsar and ternary classification of AGN-Pulsar-Other: predominantly AGN with fewer pulsars. This indicates successful ML classification attempts in the HGL region.

However, the classification of unassociated sources in the LGL region presents challenges. Due to a large number of unassociated sources in the LGL region, a single ML model tends to show peaks near the Galactic plane for different categories (such as Javier22-02). While this phenomenon can be justified for pulsars and other categories dominated by Galactic sources, it is unreasonable for extragalactic AGN. Figure \ref{figA2} left panel illustrates that the distribution of gamma-ray spectral index for unassociated sources at LGL remains unchanged (with a predominance of soft-spectrum sources, $\Gamma > 2.4$) even as the number of unassociated sources increases with longer exposure times. In previous all-sky ML classification models, individual classifiers often misclassify a large number of soft-spectrum unassociated sources as AGN candidates.

A single algorithm overlooks this situation, while multiple classifiers employing the ``All-agree" strategy classify inconsistent results as the ``MIXED" category. The bottom-right panel of Figure  \ref{figA1} shows the distribution of the MIXED class in Galactic latitude, while the left panel of Figure \ref{figA2} presents the distribution of gamma-ray spectral indices for the MIXED class. As shown in the right panel of Figure \ref{figA2}, different classifiers trained on all-sky samples yield completely different AGN-like candidate sets. By using the ``All-agree" strategy, only a small fraction of the common sample is considered as AGN-like candidates, while the majority is assigned to the ``MIXED" category. The concentration of the ``MIXED" category in the LGL region eliminates a significant number of AGN candidate sources near the Galactic plane. It also addresses the issue of an excess of soft-spectrum sources in LGL AGN-like candidates. However, it is worth noting that the ``MIXED" category is primarily concentrated near the Galactic disk. In previous attempts, the ``MIXED" category accounted for over 50$\%$ in the LGL region, meaning that over half of the sources at LGL did not receive successful classification. The nature of these sources in the ``MIXED" category still needs to be explored.

In our LGL classification framework, all training samples are LGL sources. Additionally, we have developed a BG model to handle soft-spectrum AGN. In our ML classification, the results of individual sub-classifiers tend to be consistent, providing reasonable classification results for all LGL samples.

\section{Effect of Removing Soft Spectral AGNs on the LGL Classification: Results from BG Model} \label{append:2}

In this study, we used a Bayesian-Gaussian parameter model to exclude 81 soft spectral outliers in LGL AGNs and removed them from the training and testing sets of the LGL classifier. To validate the effectiveness of the BG model in removing soft spectral AGNs, we conducted comparative experiments. These experiments involved training classification models on different datasets: Dataset 1, the complete training set without removing the 81 soft spectral AGNs; Dataset 2, the training set with 81 soft spectral outliers removed (the current dataset); and Dataset 3, where the 81 spectral outliers were add to Other-like. The results are presented in Table \ref{Tabb1}.

When training the LGL classification model using the complete Dataset 1 of 407 AGNs, 166 pulsars, and 207 Others, the balanced accuracy obtained through 5-fold stratified cross-validation was only around 75$\%$. Evaluation of 1166 unassociated LGL sources revealed approximately 600 AGN candidates, 90-168 pulsar candidates, and 400-450 Other candidates. The results indicated that AGNs dominate the unassociated LGL sources.

After removing the 81 samples identified as ``non-AGN" by the BG model from the training set, the number of predicted other-like candidates rapidly increased to approximately 700, while the number of AGN candidates decreased to around 300. The balanced accuracy of the training set improved to approximately 80$\%$, and the excess of soft-spectrum sources in AGN candidates was alleviated. The standard deviation of the cross-validation balanced accuracy decreased, indicating increased model stability. However, when re-evaluating the AGN candidates using the BG model on Dataset 2, approximately 83 sources were still classified as misclassified soft-spectrum sources, suggesting that the predicted results were still contaminated.

\cite{2022ApJS..260...53A} suggested the presence of $75\pm 4$ non-AGN spectral sources in the LGL AGN dataset, possibly originating from the Galactic component. If we include the 81 samples identified as ``non-AGN" by the BG parameter model in the other-like for model training, the number of predicted other-like candidates further increased to approximately 850, while the number of AGN candidates decreased to around 200 (see Table \ref{Tabb1}). The number of AGN-like candidates is similar to the high-confidence candidates obtained by BG model re-evaluating of Dataset 2 ML classification model. This suggests that these 81 soft-spectrum sources may indeed belong to a category other-like than AGN and pulsars, significantly impacting the ML classification of LGL sources. However, further evidence is still needed to support this viewpoint.

There are a total of 781 associated LGL sources, and the only difference between these three training sets is the presence of 81 AGN samples. The influence of different training sets on the accuracy of ML models on training and testing sets is limited (approximately 5$\%$), but it leads to significant differences in the prediction number for unassociated sources. Different training sets yield distinct results regarding whether low-latitude unassociated sources are dominated by AGNs or Other-like sources. This highlights the significant impact in prediction of even minor changes in the dataset during the ML model training process.

Although the prediction results for AGN and Other categories remain relatively stable after removing low-confidence soft-spectrum AGNs, there are still significant differences in the number of pulsar candidates among different classifiers. This raises concerns about the reliability of pulsar candidate results and may require further consideration of the purity of pulsar training samples.

\begin{table}
\centering
\caption{Classification Results of Low Galactic Latitude Sources with Different Trainingsets}\label{Tabb1}
\hspace{-0.9cm}
\resizebox{0.49\textwidth}{!}{
\begin{tabular}{ccccc}
\hline \hline
\multirow{2}{*}{Estimator}&\multirow{2}{*}{Test balanced accuracy}&\multirow{2}{*}{AGN}&\multirow{2}{*}{PSR}	&\multirow{2}{*}{OTHER}\\
\\
\normalsize(1) & \normalsize(2) & \normalsize(3) &\normalsize(4)  &\normalsize(5)\\
\hline
LR (Dataset1)	&$0.754\pm 0.049$&563&151&452\\
SVM (Dataset1) &	$0.764\pm 0.044$ &603&131&432\\
RF (Dataset1) &$0.760\pm 0.041$&634&90&442\\
MLP (Dataset1)&$0.754\pm 0.041$& 593&168&405\\
\hline
LR (Dataset2)	&$0.809\pm 0.021$&318&136&712\\
SVM (Dataset2) &	$0.797\pm 0.039$ &306&155&706\\
RF (Dataset2) &$0.802\pm 0.021$&300&106&760\\
MLP (Dataset2)&$0.788\pm 0.046$&260&199&707\\
\hline
LR (Dataset3)	&$0.817\pm0.024 $&202&103&861\\
SVM (Dataset3) &	$0.797\pm0.025$ &214&115&837\\
RF (Dataset3) &$0.811\pm 0.024$&218&67&881\\
MLP (Dataset3)&$0.801\pm0.030 $&230&189&747\\
\hline
\end{tabular}}\\
{\footnotesize{{Note.  Column (1) represents the estimator, which refers to the classification algorithm used. Column (2) shows the test balanced accuracy, which represents the accuracy of the model on the test set. Column (3)- (5) indicates the number of sources classified as AGN-like, pulsar-like and other-like. Dataset1, Dataset2, and Dataset3 refer to different datasets used for model training and test.  }}}
\end{table}

\section{Effect of Hyper-parameter Tuning and Oversampling Methods on the HGL Classification} \label{append:3}

The HGL AGN-like and pulsar-like binary classification is a typical imbalanced classification problem. In the scikit-learn library, many classification algorithms have a hyper-parameter called ``class$\_$weight" \citep{scikit-learn}. This hyper-parameter is used to adjust the weights of different class samples in imbalanced classification problems, to avoid bias towards the majority class. The default value for  ``class$\_$weight" is ``None", which means no artificial weights are added during the training process. For a binary classification problem where class $a$ has $N_a$ samples and class $b$ has $N_b$ samples, setting the  ``class$\_$weight" to ``balanced" assigns a weight of $N_b / (N_a + N_b)$ to class $a$ and a weight of $N_a / (N_a + N_b)$ to class$ $b. This way, the majority class is suppressed while the minority class is emphasized, achieving balanced classification.

In this study, four supervised ML classification algorithms were used, with the MLP algorithm not having a ``class$\_$weight" hyper-parameter. For the other three algorithms, LR, SVM, and RF, the ``class$\_$weight" was set to ``balanced," and they were compared with the classifiers that did not use the ```class$\_$weight" parameter. After training and optimizing the models using the methods described in Section \ref{sec:classifier}, the balanced accuracy of these seven classifiers was evaluated on the 5-folds stratified cross validation, and predictions were provided for the 1125 unassociated sources (see Table \ref{Tabc1}).

SMOTE is a synthetic oversampling method used to address class imbalance classification issues \citep{chawla2002smote}. In imbalance  classification problems, the minority class has a smaller number of samples, leading to poor performance of the classifier in learning and predicting the minority class. SMOTE balances the dataset by synthesizing new samples for the minority class, improving the performance of the classifier. In this study, the SMOTE algorithm from the imbalance-learn library was used to oversample the pulsar samples to match the number of AGN samples \citep{lemaitre2017imbalanced}. The constructed dataset was then used to train LR, SVM, RF, and MLP classifiers for HGL classification. After training and optimizing these classifiers, their performance on the test set and the prediction results were evaluated and presented (see Table \ref{Tabc1}).

In this context, balanced accuracy was used instead of accuracy to evaluate the models. As shown in the Table \ref{Tabc1}, increasing the weight of the pulsar samples or increasing their number through SMOTE led to an improvement in the accuracy of the pulsar class, resulting in an overall increase in balanced accuracy. By adjusting the weight parameters by turning hyper-parameter, the balanced accuracy on the test set increased from 90$\%$ to over 96$\%$. However, there were significant differences in the predictions between different classifiers, and the predicted number of pulsar candidates also showed large fluctuations. After applying SMOTE for weight adjustment, the balanced accuracy on the test set rapidly increased from 90$\%$ to over 99$\%$. Different classification algorithms responded differently to SMOTE, for example, MLP showed little change in the prediction results for the 1125 unassociated sources, while the other three algorithms exhibited significant changes. Similarly, there were significant differences in the predictions between different classifiers, and the predicted number of pulsar candidates also showed large fluctuations.

Although these methods significantly improve the accuracy on the training and testing sets, they also widen the differences in prediction results among different classifiers. Therefore, it cannot be concluded that the classifiers have been optimized and resulted in more reliable predictions.

These results indicate that artificially changing the weights of samples or using oversampling methods like SMOTE does not necessarily lower the performance of classifiers on the training and test samples; in fact, they may even enhance the performance. However, they significantly alter the prediction results. This reminds us that the inherent differences in sample quantities are important parameters in machine learning training. It also highlights that making subtle changes to the dataset during the model training process can have an impact on performance metrics such as accuracy and may lead to significant differences in classification results.

\begin{table}
\centering
\caption{The classification results of the high Galactic latitude classifier using different class weights }\label{Tabc1}
\hspace{-0.9cm}
\resizebox{0.49\textwidth}{!}{
\begin{tabular}{ccccc}
\hline \hline
\multirow{2}{*}{Estimator}&\multirow{2}{*}{Test balanced accuracy}&\multirow{2}{*}{AGN}&\multirow{2}{*}{PSR}	\\
\\
\normalsize(1) & \normalsize(2) & \normalsize(3) &\normalsize(4)  \\
\hline
LR (class$\_$weight =None)	&$0.884\pm 0.027$&1042&83\\
SVM (class$\_$weight =None) &	$0.907\pm 0.034$ &1055&70\\
RF (class$\_$weight =None) &$0.898\pm 0.026$&1059&66\\
MLP (class$\_$weight =None)&$0.926\pm 0.028$& 1032&93\\
\hline
LR (class$\_$weight =balanced)	&$0.961\pm 0.011$&846&279\\
SVM (class$\_$weight =balanced) &	$0.965\pm 0.011$ &802&323\\
RF (class$\_$weight =balanced) &$0.953\pm 0.021$&959&166\\
\hline
LR (SMOTE)	&$0.976\pm0.022 $&869&256\\
SVM (SMOTE) &	$0.982\pm0.023$ &859&266\\
RF (SMOTE) &$0.995\pm 0.015$&1000&125\\
MLP (SMOTE)&$0.997\pm0.013$&1039&86\\
\hline
\end{tabular}}\\
{\footnotesize{{Note.  Column (1) represents the estimator, which refers to the classification algorithm used. Column (2) shows the test balanced accuracy, which represents the accuracy of the model on the test set. Column (3) - (4) indicates the number of sources classified as AGN-like and pulsar-like.}}}
\end{table}

\label{lastpage}
\end{document}